\documentclass[aps,pra,reprint,showpacs,twocolumn,superscriptaddress,eqsecnum]{revtex4-2}

\usepackage[margin=0.75in]{geometry}
\usepackage{graphicx}
\usepackage{amsmath,amssymb}
\usepackage[caption=false]{subfig} 
\usepackage{xcolor}
\usepackage{hyperref}
\usepackage{titlesec}
\usepackage{comment}
\usepackage{makecell}
\usepackage{booktabs}
\usepackage{siunitx}
\usepackage{multirow,array}
\usepackage[breakable,skins]{tcolorbox}
\usepackage{amsthm}

\makeatletter
\newenvironment{subfigure}[2][b]{%
  \begin{minipage}{#2}%
    \def\caption##1{\gdef\@subcap{\parbox{\linewidth}{\centering ##1}}}%
    \gdef\@subcap{}
}{%
    \ifx\@subcap\@empty\else
      \par\vspace{4pt}{\small\@subcap}
    \fi
  \end{minipage}%
}
\makeatother

\makeatletter
\@ifundefined{widetext}{%
  
}{}
\makeatother

\hypersetup{
    colorlinks,
    linkcolor={red!50!black},
    citecolor={blue!50!black},
    urlcolor={blue!80!black}
}

\newcommand{\kt}{\kappa(t)}
\newcommand{\tcoll}{\tau_{\mathrm{coll}}}

\newcommand{\Abar}{\bar{A}}
\titleformat{\section}{\large\bfseries}{\thesection}{1em}{}
\newcommand{\chapter}[1]{\section{#1}}
\date{}

\theoremstyle{remark}
\newtheorem{remark}{Remark}
\newcommand{\kss}{\kappa}
\newcommand{\kinit}{\kappa(0)}

\begin{document}
\title{Non-Maxwellian Velocity Statistics in Supercooled Liquids and Their Possible Relation to Super-Arrhenius Viscosity}

\author{Giorgi Tsereteli}
\email{g.tsereteli@wustl.edu}
\affiliation{Department of Physics, Washington University, St.~Louis, Missouri 63130, USA}
\author{Zohar Nussinov}
\email{zohar@wustl.edu}
\affiliation{Department of Physics, Washington University, St.~Louis, Missouri 63130, USA}
\affiliation{LPTMC, CNRS-UMR 7600, Sorbonne Universit\'e, 4 Place Jussieu, 75252 Paris cedex 05, France}
\affiliation{Rudolf Peierls Centre for Theoretical Physics, University of Oxford, Oxford OX1 3PU, United Kingdom}
\date{\today}

\begin{abstract}
For particles of fixed mass, classical equilibrium statistical mechanics dictates a Maxwellian velocity distribution determined solely by the temperature, regardless of the interactions, density, or structure. Supercooled glass forming liquids realize long lived metastable states that evade equilibrium crystallization and may thus violate assumptions underlying Maxwellian statistics. We numerically demonstrate that supercooled liquids can exhibit persistent non-Maxwellian velocity distributions with deviations connected to their exceptionally slow super-Arrhenius relaxation. Our work is motivated by a general result establishing that long lived metastable states may exhibit finite width distributions of intensive variables. A distribution of temperatures implies non-Maxwellian velocity statistics. We test this prediction by introducing stochastic thermostats that generate stationary states while, unlike conventional thermostats, not imposing Maxwellian velocity distributions. Simulations with these thermostats yield long lived states that have, by comparison to Maxwellian velocity distributions, an excess kurtosis $0<\kappa\lesssim0.3$. Crystallization is strongly impeded with increasing $\kappa$. In a minimal description, temperature fluctuations are characterized by a dimensionless width $\overline{A}$ with $\kappa\simeq3\overline{A}^{2}$. The nearly constant $\overline{A}$ (of an average value $0.08$ and standard deviation $0.03$) found in viscosity data collapse across $45$ glass formers and in specific heat signatures is consistent with kurtosis found in our simulations. Long time non-Maxwellian velocity statistics may thus link slow relaxation, transport, and thermodynamic measurements. Independent of the tested theory, the stochastic thermostats that we introduce offer a molecular dynamics route to non-Maxwellian velocity statistics.
\end{abstract}

\maketitle

\section{Thermostats and velocity distributions}
\label{cpt1}

\subsection{Theoretical Introduction}

Supercooled liquids realize an enigmatic long-lived metastable state \cite{doi:10.1021/jp953538d,bib:angell,BB} of a lifetime that is $10^{10}$ to $10^{17}$ times longer than their natural underlying microscopic timescales. On general grounds, in the limit of divergent metastable state lifetimes, all local few particle expectation values \cite{Nussinov2024} (see also Appendix \ref{app:theorem}) can be expressed as unique weighted averages of their equilibrium counterparts. In the current work, we focus on a corollary of this result that affords {\it a link between the deviation of the single particle velocity from the equilibrium Maxwellian distribution to the shear viscosity}. In the current numerical work, we test this predicted link with experiment to find consistent agreement. We will largely quantify the latter deviation via the excess kurtosis of the velocity distribution relative to that in equilibrium. We first outline theoretical elements underlying this suggested link between the single particle velocity statistics and the shear viscosity in a supercooled liquid:

(i)~For a classical particle of fixed mass, regardless of details of the system (include specific interparticle interactions), the equilibrium velocity distribution is Maxwellian and is set by the temperature alone.

(ii)~Although temperature is strictly an equilibrium concept, a corollary of Ref. \cite{Nussinov2024} implies that the velocity distribution of a long lived metastable state at a given empirically determined temperature $T$ can be expressed as a weighted sum of Maxwellians of effective temperatures $T'$ drawn from a distribution $P(T|T')$. A finite width of this distribution $P$ yields a non-Maxwellian velocity distribution.

(iii)~This width is captured by a dimensionless parameter $\overline{A}:=\sigma_{T'}/\langle T'\rangle = \sigma_{T'}/T$, which is related to the measured excess velocity kurtosis $\kappa$ by 
\begin{eqnarray}
\label{eq:kappa_Abar} 
\kappa\simeq3\overline{A}^{2}.
\end{eqnarray}

(iv) $\overline{A}$ determines the weight of the distribution $P$ that lies above the equilibrium liquidus temperature (the temperature above which the system is completely liquid). In equilibrium, only states of temperature $T'$ above the liquidus temperature contribute to sustained hydrodynamic flow. Thus, the shear viscosity $\eta(T)$ is controlled by this statistical weight above the liquidus temperature. Putting all of the pieces together provides the said connection between velocity statistics and the temperature dependence of the viscosity.

The resulting expression for $\eta(T)$ depends only on the liquidus temperature and $\overline{A}$ \cite{Nussinov2024,grant16,grant15,preprint2}. As we will illustrate, this expression does not introduce additional conjectured temperatures of the type found in the Vogel Fulcher Tammann (VFT) law ~\cite{bib:vft,bib:vft2,bib:vft3}. When compared against experimental data of $45$ glass formers, the expression is consistent over sixteen decades of the viscosity. Statistical comparisons ~\cite{grant15,preprint2} further indicate that the resulting single parameter form describes the data more accurately than the three parameter VFT expression. 

Undertaking tests, in conventional numerical experiments, the predicted link between velocity statistics and the viscosity encounters some hurdles. Conventional MD thermostats enforce the equilibrium Maxwell velocity distribution and therefore preclude, from the very start, the observation of possible deviations from Maxwellian statistics. We consequently employ alternative dynamical schemes that leave the velocity distribution unconstrained. We introduce and employ stochastic thermostats that do not impose a Maxwellian velocity distribution. Technical details are provided in Appendix~\ref{app:thermostats}. These thermostats allow the velocity distribution to be measured rather than prescribed. With the aid of these thermostats, we determine the excess kurtosis and the associated value of $\overline{A}$ and compare it with that inferred from viscosity data. If the excess kurtosis were merely a transient consequence of coupling to the bath, it would be expected to disappear rapidly. In simple solvable instances, uniform  Boltzmann equation considerations indeed suggest a collision rate dominated relaxation toward a Maxwellian distribution (Appendix~\ref{app:boltzmann}). The finite excess kurtosis probability distributions that aim to explore allow for both nontrivial spatial and momentum distributions and are trivially time translationally invariant. 

The excess kurtosis is also related to other properties of the metastable state. The probability of crystallization decreases once $\kappa\gtrsim0.15$. Related signatures appear in the distribution of single particle potential energies. The excess kurtosis is therefore associated with properties of the metastable state beyond the velocity distribution itself.

Before proceeding further, we briefly regress to a broad introduction of the general problem at hand- that of supercooled liquids and the glass transition. Notwithstanding extensive progress that resulted from intense investigations over many decades  \cite{doi:10.1021/jp953538d,ediger2000spatially,Debenedetti2001}, the dynamics of supercooled glass-forming fluids near their glass transition remain ill-understood. When simple single substance liquids are cooled slowly, they typically crystallize below their freezing temperature (or melting temperature $T_{m}$). Similarly, in numerous complex metallic alloys, instead of a single melting temperature, the equilibrium system is purely liquid above the above discussed liquidus temperature $T_{l}$ and purely solid below the so, called solidus temperature; for simple single component systems, the liquidus and melting temperatures are one and the same, $T_{l} = T_{m}$. In the opposite limit---rapid cooling, or ``supercooling''---crystallization is kinetically bypassed, and the system enters a long lived metastable state. After a material-dependent waiting time (but before the system eventually crystallizes or ``devitrifies''), rich universal phenomena emerge: local observables and their dynamics converge to a steady state that remains poorly understood to this day.  Remarkably, despite their vastly different properties, the supercooled liquid and the equilibrated crystalline phase are governed by the same underlying  Hamiltonian. Adding further complexity, many glassformers are  multicomponent systems in which equilibrium melting is not a sharp transition. Instead, it occurs over a temperature range bounded by the solidus temperature where melting begins and a higher temperature liquidus temperature $T_{l}$ denotes where it is completed.

At sufficiently low temperatures ($T < T_g$), supercooled liquids form, on typical experimental time scales, amorphous solids---glasses. This time scale definition of $T_g$ (relying on typical time scales rather than conventional markers of a phase transition) is somewhat ad-hoc. The ``glass transition'' at temperature $T=T_{g}$ has been known empirically for millennia and is  the focus of intense modern research. This ``transition''  differs sharply from conventional thermodynamic transitions. Most strikingly, relaxation times can increase by a factor of $10^{14}$ as temperature drops, with no corresponding signature in thermodynamic quantities such as the specific heat. The most celebrated description of this dramatic dynamical slowdown is the VFT law~\cite{bib:vft,bib:vft2,bib:vft3}
which predicts a divergence of viscosity and relaxation times at finite temperature,
\begin{equation}
\eta(T) = \eta_0 \exp\!\left[\frac{D\,T_0}{T - T_0}\right].
\label{eq:VFT_law}
\end{equation}
Here, \( \eta_0 \), \( D \), and \( T_0 \) are liquid dependent parameters. Regardless of the veracity of this form, including notably whether a true transition appears at a temperature $T_0$ of whether the latter is only an artifact when attempting to fit with the VFT equation, the dynamics of supercooled liquids depart much more radically from Arrhenius activated dynamics than those in conventional fluids \cite{jing}. The extreme (``super-Arrhenius'') slowing down of liquid dynamics as their temperature is lowered is a hallmark of supercooled liquids. If Eq. (\ref{eq:VFT_law}) were to be taken literally, the temperature $T_{0}$ would mark an essential singularity at which the relaxation times  scales diverge; this is the so-called ``ideal glass'' temperature that many theories attempt to rationalize. The VFT law and related fits predicting a true divergence of relaxation times at temperatures below the glass transition have been called into question by experiment~\cite{amber2013,grant3}. Nonetheless, the VFT fit is well known to capture the quintessential behavior of supercooled liquids above their glass transition temperature. Regardless of the accuracy of specific fitting forms, the rapid increase of the viscosity of these liquids (especially in so-called ``fragile'' glassformers \cite{bib:angell})  as their temperature drops is a hallmark of these systems.

A vexing problem in understanding these systems are their very slow dynamics (these slow dynamics are typically especially acute in simulated {\it in silico} systems that aim to provide a microscopic picture of what occurs). Numerical studies using MD simulations in both two and three dimensions have provided valuable information on relaxation behavior \cite{Flenner2015, Illing2017}, but a unified theoretical description remains elusive. Several theoretical approaches have been developed to address this problem. These can largely be divided into those focusing on dynamic versus thermodynamic descriptions. The former approaches include the celebrated mode coupling theories \cite{bib:mct1,bib:mct2,bib:mct3,mode_coupling_L} and ideas concerning ``dynamic facilitation'' \cite{Mirigian2013unified, Palmer1984,Berthier2021,keys} seeking to describe heterogeneous relaxation through constrained correlated mobility events. There are numerous statistical mechanics approaches, e.g., \cite{AGlass,Goldstein2,RFOT:crit_assess,BB,Ted87b,ktw,ref:lubchenkowolynes,bib:ag2,bookZamponi,ref:angelaniPEL,ref:parisiglasstrans,Patrick2017,Steve_Gilles,bib:KKZNT1,bib:KKZNT3} that assume that some core ideas from equilibrium statistical mechanics may still remain applicable for the long lived (yet still metastable) supercooled fluid state.
These thermodynamic approaches often build on mean-field concepts, compare the free energies of the long lived metastable supercooled state of fluids below their melting temperature to those of the bona fide equilibrium solid (crystalline) state, discuss energy landscapes, configurational entropy effects, avoided transitions, and explore countless other interesting ideas.
Notable work has also modeled localized mobility defects using statistical mechanical approaches to capture slow dynamics \cite{Speck2019}.

For decades, it has been unquestionably assumed that the particle velocity distribution in a supercooled liquid follows the Maxwell-Boltzmann form, yet this has not been tested to date. The prevailing view is that the many body system navigates a complex high-dimensional energy landscape spanned by generalized coordinates specifying atomic configurations, while the momentum distribution remains a simple Maxwell-Boltzmann form.

This assumption has persisted even after the discovery of spatially nonuniform ``dynamical heterogeneities'' \cite{DH1,DH2,DH3,DH4,DH5,dh10,volynes,dh11,dh12}. Within configuration energy landscape approaches \cite{BB,Goldstein2}, a Maxwellian velocity distribution is still assumed for supercooled liquids---the effective potential may differ from that in the equilibrium system, but the kinetic energy is taken to be that of an equilibrium system at fixed temperature.

The general theorem of Ref. ~\cite{Nussinov2024} establishes that the reduced few body probability density of any stationary state of an open system is a weighted superposition of equilibrium densities at different effective values of all state variables. This result is general: it makes no assumption about how the stationary state was prepared or what broadened its distribution of effective state variables. In the present work, we focus on the distribution $P(T|T')$ of effective local temperatures. This is natural since, for particles of fixed mass, the single particle equilibrium velocity distribution is Maxwell-Boltzmann at temperature $T'$, independent of the interaction potential, the particle density, and all other state variables. (More generally, the distribution extends to all state variables including the local particle number density.) One mechanism by which $P(T|T')$ acquires a finite width is the following~\cite{preprint6} (Appendix \ref{rem:drive}). When a liquid is rapidly cooled, the external environment must couple to an extensive fraction of all atomic sites in order to change the energy density at a finite rate. The fluctuations of this common environment are imparted simultaneously to spatially distant sites, generating long range correlations and a system size-independent ($\mathcal{O}(1)$) standard deviation $\sigma_\epsilon$ of the global average energy density. Once cooling ceases, local uncorrelated noise destroys the long range covariance between sites, so the global average energy becomes self-averaging with fluctuations of order $1/\sqrt{N}\to 0$. However, the \emph{spatial} heterogeneity that was imprinted across sites during the drive --- the $\mathcal{O}(1)$-wide distribution of effective local temperatures $\{T'_i\}$ across different spatial regions --- cannot be removed: the system is trapped by nucleation barriers and structural arrest, so rapid supercooling prevents global equilibration and only local rearrangements on scales of $\sim\mathsf{n}$ particles are possible. It is the width of $T'$ in the \emph{reduced few body local probability density} --- the spatial variation of effective local temperatures across the system --- that constitutes the finite-width $P(T|T')$ of the central theorem and gives rise to non-Maxwellian velocity statistics. A detailed derivation by explicit integration over the external drive is given in Appendix \ref{app:theorem}. The system then relaxes toward the nearest stationary state it can reach. True equilibrium, the crystal, is inaccessible: nucleation is suppressed by a large free-energy barrier near the melting temperature, and structural relaxation is arrested by the high viscosity at low temperatures. The system instead settles into the long lived metastable supercooled state, stationary on all timescales satisfying $\tau_{\rm micro} \ll t \ll \tau_{\rm xtal}$. The broadening of $P(T|T')$ is essential: if it were sharp, the supercooled liquid would be indistinguishable from a true equilibrium state of the same Hamiltonian, which is the crystal. Earlier work~\cite{book_TOP,grant16,grant15,preprint2} developed this physical picture and its consequences for viscosity and other observables of supercooled liquids, invoking a notion of \emph{global} stationarity: the long time average of the full many body density was assumed to be stationary. The key advance of Ref.~\cite{Nussinov2024} is to invoke instead \emph{local} stationarity. Indeed, empirically, the long time average of local few body observables --- and hence of the reduced few body probability density --- equals the instantaneous local reduced probability density in the metastable supercooled state. This weaker and physically more natural condition is what allows the theorem to hold for a metastable system that is not globally stationary. The velocity distribution is related to $P(T|T')$ through a Laplace-transform relation. In practice, we do not attempt to invert this (ill-posed) transform but extract only its width via the moment relation Eq.~(\ref{eq:kappa_Abar}). The width parameter $\overline{A}$ of this distribution is the same quantity that collapses the viscosity of 45 experimental glass-forming liquids over sixteen decades onto a single universal curve~\cite{grant15,Nussinov2024}. Conventional thermostats, by enforcing Maxwell-Boltzmann velocity statistics, erase this signal by construction. The thermostats introduced in the present work do not.

A central aim of our work  is to \textbf{systematically study the velocity distributions} in supercooled liquids via MD simulations. Thermostats used in most conventional simulations (e.g., the Nos\'e-Hoover \cite{Nose,Hoover}, Andersen \cite{Andersen_thermostat}, and Langevin \cite{Langevin1,Langevin2} thermostats) are all guaranteed, by construction, to yield a Maxwell-Boltzmann velocity distribution. Thus, due to the nature of these thermostats, deviations from the equilibrium Maxwellian velocity distribution may not be observed numerically when these conventional thermostats are used. Our MD results demonstrate that, when these thermostats are \emph{not used}, supercooled Kob-Andersen type \cite{Kob-Andersen} systems exhibit a \emph{non-Maxwellian velocity distribution}.

Prior to the current work, dynamical heterogeneities have been numerically probed and assumed to occur physically in the form of spatially fluctuating relaxation rates. Theoretically, much effort has been devoted to study how dynamic facilitation affects supercooled liquids in MD simulations. One approach \cite{PhysRevLett.132.258201} creates spatial variation by applying rapid swap Monte Carlo dynamics to specific regions while using a global Nos\'e--Hoover thermostat to maintain equilibrium.
The method that we describe in this work differs in that we do not preselect fast regions but rather aim to simulate the system as it is cooled in real experiment without imposing a Maxwellian velocity distribution. We modify the thermostat so that stronger coupling appears randomly in both space and time. This generates dynamically heterogeneous regions naturally, letting us compare the resulting behavior with that of standard thermostats at the same temperature.

For completeness, we briefly note related work. Recent work \cite{gao2024unifiedpercolationscenarioalpha} has introduced a dual-temperature percolation framework for deeply supercooled glass formers, proposing that a single equilibrium temperature may not fully capture inherently non-equilibrium dynamics. Similarly, another study \cite{fast} found that the subset of particles in truly equilibrated (i.e., non-supercooled) liquids that transit over distances longer than average can deviate from Maxwellian behavior. Previous studies \cite{Whitmer2010} showed that colloidal systems with different boundary conditions can develop non-Maxwellian velocity distributions. Signatures of nontrivial low temperatures dynamics were investigated in  \cite{OriginWings,WingExplained}; in these works, asymmetric features in relaxation spectra were observed and subsequently rationalized. In Ref. \cite{EmergentFaci}, dynamic facilitation was proposed to emerge from localized stress excitations modeled using Markovian dynamics. Our work shares this motivation but takes a more physically grounded approach. As alluded to above, instead of introducing explicit stress-based rules, we probed the dynamics of supercooled fluids using a custom stochastic thermostat designed to mimic physical cooling processes. Specifically, the thermostat simulates random central collisions between system particles and fictitious bath particles drawn from an equilibrated colder reservoir. This creates local energy exchange through collision-like events, similar to how real hot systems may cool down. Initially, there is a large difference in velocities between the hot and cold particles. As they equilibrate, the distribution of particle velocities become less broad yet might still (and indeed does) self consistently stabilize towards a broader than
a Maxwellian distribution. The distribution emulates the non-uniform cooling background. The broader the background distribution from which the velocities are drawn from stochastically in our collision based framework, the broader the single particle velocity distribution of the system. This collision-based energy exchange naturally produces dynamically heterogeneous regions in space and time. We reiterate that the resulting dynamics that we find indeed differ qualitatively from those generated by conventional global thermostats such as Nos\'e--Hoover or Berendsen, which impose equilibrium constraints without modeling physically local cooling mechanisms. Recent experimental studies of metallic glass formers using temperature scanning X-ray photon correlation spectroscopy \cite{frey2024interplayliquidlikestressdrivendynamics} support the idea that multiple distinct dynamical processes coexist as systems approach glassy states. Earlier computational work \cite{doi:10.1021/acs.jpcb.5b08912} demonstrated that dynamic heterogeneities emerge even in minimal MD models such as two-dimensional Lennard-Jones liquids, which are among the systems we examine here. Additional insight comes from \cite{lam2024distinguishableparticleglassycrystalsimplest}, which shows that glassy dynamics in three-dimensional Lennard-Jones systems can be revealed through potential energy distributions rather than solely through structural or kinetic observables. We similarly find that local potential energy distributions provide a robust probe of supercooled dynamics, as they fluctuate more slowly during relaxation than velocity distributions or structural measures. Recent work has used machine learning to predict dynamical quantities such as four-point correlation functions and local temperature fields \cite{Jung2023Predicting}. In constructing our MD systems, we followed established methods and benchmarks  \cite{LJisotherm,LJmelting,MDnuc,Hafskjold_2019,visc,Zhao2008BriefIT,https://doi.org/10.12921/cmst.2024.0000005}.

Building on the above noted phase space concepts and the  more general, theory unbiased, quest to understand dynamics in supercooled liquids that are unconstrained by conventional thermostats that automatically  impose a Maxwell-Boltzman distribution of velocities, we studied the dynamics in these systems with physically motivated thermostats.

The prevailing interpretation of dynamical heterogeneity assumes that notwithstanding the existence of spatially varying dynamics, the velocities themselves maintain a Boltzmann distribution throughout. In the present work we question this assumption. If spatially varying regions harbor genuinely faster/slower particles, then the single particle velocity distribution may be broader than that in Maxwellian systems and the respective excess kurtosis $\kappa$ will not vanish.

We study both two- and three-dimensional systems with a dominant focus on the latter. Of all the figures in this paper, only Fig.~\ref{fig:2Dkurt} reports our results on two-dimensional systems. In three dimensions, the thermostat's role as the stabilizer of the system becomes apparent much more quickly for comparably sized systems, so the thermostat-present versus thermostat-absent comparison is not feasible. Instead, we compare the same system evolved with two markedly different thermostats while modulating the degree of non-equilibrium through the velocity kurtosis. The advantage of the three-dimensional systems is the naturally emergent crystallization at a consistent melting temperature, unlike the two-dimensional case where Mermin--Wagner fluctuations preclude such an outcome.

\subsection{Thermostat Mechanisms and Non-Maxwellian Sampling}

To investigate how thermostats influence supercooled molecular dynamics, we design cooling protocols where different thermostats cool an initially equilibrated liquid to the same target temperature using comparable cooling schedules. All thermodynamic parameters remain constant; only the thermostat coupling is varied. This isolates the effect of the realistic coupling needed to achieve rapid cooling on the emergent dynamics.

We consider two broad classes of thermostats: Continuous coupling schemes, such as Langevin, Berendsen, and Nos\'e--Hoover thermostats, act on all particles at every timestep via weak, persistent interactions that continuously regulate the kinetic degrees of freedom. In contrast, collision-based stochastic schemes, including the Heavy-Tailed Lowe--Andersen (HTLA) and Two-Bath Andersen thermostats, operate through discrete, instantaneous perturbations applied to a randomly selected subset of particles, resulting in intermittent but typically stronger modifications of particle velocities. Details concerning these schemes, both the conventional thermostats together and the two stochastic thermostats introduced in this work, are provided in Appendix~\ref{app:thermostats}. 

Continuous thermostats maintain near Maxwellian velocity statistics through frequent small adjustments. In contrast, collision-based stochastic thermostats introduce intermittent, localized energy exchanges that broaden the instantaneous velocity distribution even when the mean kinetic energy (temperature) remains the same.

\begin{figure}
    \centering
        \centering
        \includegraphics[width=1\linewidth]{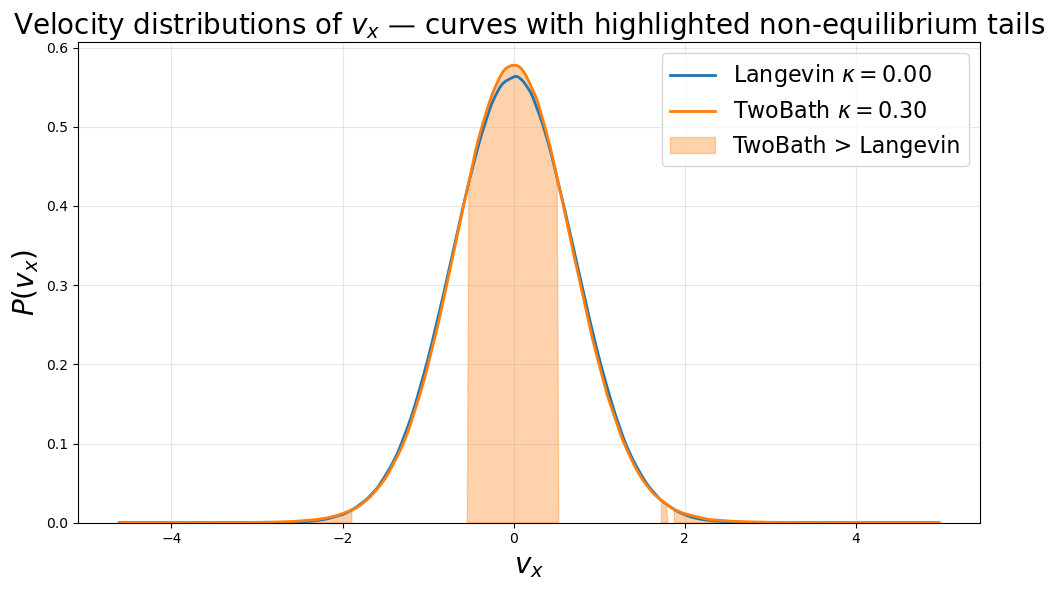}
        \label{fig:vel_stoch}
    \caption[Velocity Distributions]{ Comparison of single particle velocity distributions (for a Cartesian component $v_x$)
    in the supercooled fluid  obtained using different thermostats with a Maxwellian (Gaussian) velocity distribution of the same mean (zero) and standard deviation (set by the temperature). The orange curve corresponds to the Two baths Andersen thermostat, which exhibits enhanced excess kurtosis ($\kappa = 0.30$) relative to the Maxwellian reference. The blue curve shows the distribution obtained using the Langevin thermostat,
    which closely follows the Maxwellian form ($\kappa \approx 0$). The highlighted regions indicate where the supercooled state distribution satisfies
$P_{\mathrm{supercooled}} (v_x)> P_{\sf eq.}(v_x)$
featuring heavier velocity tails at the same nominal temperature. Each distribution was obtained from a sample of 15000 velocities.
    }
\label{fig:velocity_comparison}
\end{figure}

These differences are evident in Fig.~\ref{fig:velocity_comparison},
where normalized velocity distributions are compared at equal temperatures.
The stochastic thermostat produces visibly heavier tails, whereas the
Berendsen case closely follows the Maxwellian form.

\begin{figure*}[t]
    \centering
    \begin{subfigure}{0.48\textwidth}
        \centering
        \includegraphics[width=\linewidth]{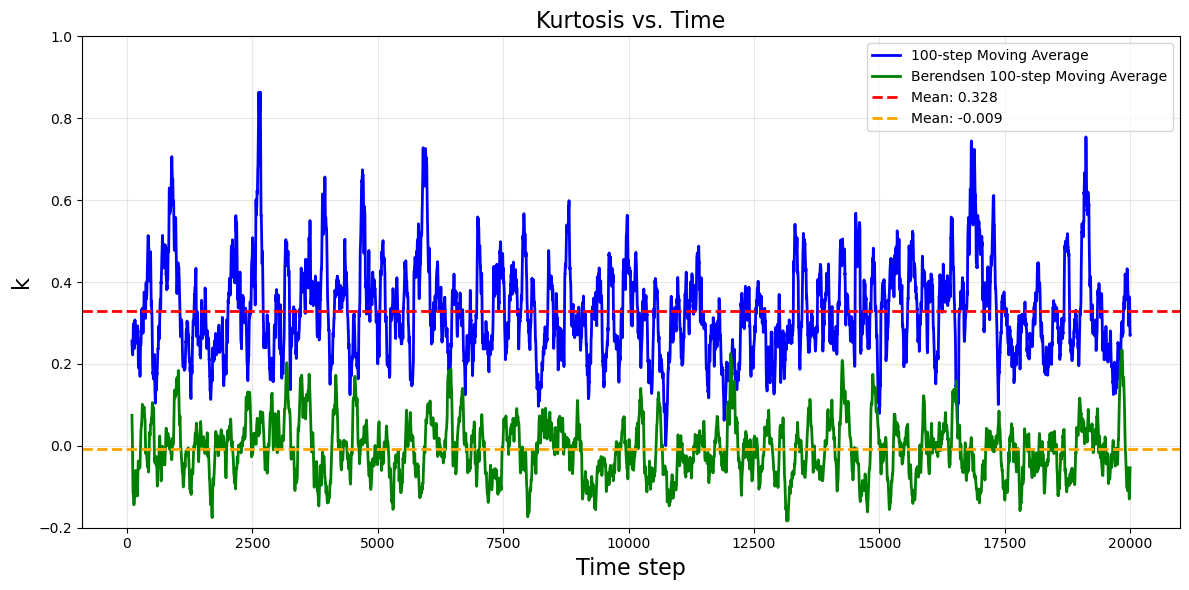}
        \caption{\scriptsize Kurtosis comparison}
        \label{fig:K_langevin}
    \end{subfigure}
    \hfill
    \begin{subfigure}{0.48\textwidth}
        \centering
        \includegraphics[width=\linewidth]{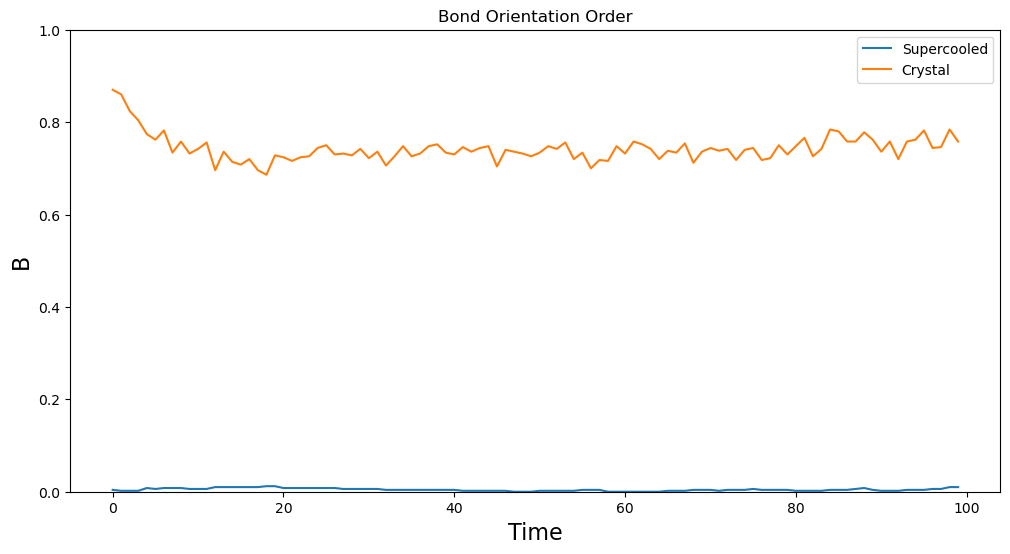}
        \caption{\scriptsize Bond order comparison}
        \label{fig:K_twobath}
    \end{subfigure}
    \caption{
    Combined structural and dynamical comparison between equilibrium crystallization (Langevin dynamics) and nonequilibrium supercooling (Two-Bath Andersen thermostat).
    (a) Velocity kurtosis during relaxation: the Two-Bath thermostat produces a noisy steady state with higher excess kurtosis (\(\kappa = 0.32 > 0.004\)), while Langevin yields smooth equilibration and crystallization. The curves are shifted so the Gaussian value (\(k=3\)) maps to zero.
    (b) Structural evolution: the bond orientational order parameter $B$ distinguishes the supercooled state (blue) from the crystalline state (orange). Both systems approach stationary regimes as seen from saturation of the self-intermediate scattering function.
    }
    \label{fig:combined_kurtosis_structure}
\end{figure*}

The dynamical consequences are summarized in
Fig.~\ref{fig:combined_kurtosis_structure}. Langevin dynamics promote smooth equilibration and crystallization, while the Two-Bath stochastic
thermostat drives the system into a noisy steady state with sustained
deviations from Gaussian velocity statistics.

\subsection{Velocity Kurtosis as a Measure of Non-Equilibrium Sampling}

To quantify deviations from Maxwellian statistics, we compute the
kurtosis of a Cartesian velocity component,
\begin{equation}
k := \frac{\langle (v_x-\langle v_x \rangle)^4 \rangle}{\langle (v_x  - \langle v_x \rangle)^2 \rangle},
\end{equation}
with the average $\langle \cdot \rangle$ performed over all particles in the system within the ensemble that we simulate. In the absence of externally imposed symmetry breaking (such as in the supercooled liquids that we simulate), $\langle v_x \rangle =0$. Independent of temperature and spatial dimensionality, for a Maxwellian (a Gaussian distribution in each Cartesian velocity component),
the kurtosis is, trivially,
$k = 3$.
Although all of our simulations are classical, we note for completeness that the same value $k=3$ trivially appears for the Gaussian momentum distribution in the quantum arena for finite temperature harmonic atomic vibrations. To probe the deviations from a Maxwell-Boltzmann distribution, we therefore focus on the excess kurtosis
\begin{equation}
\kappa(t) := k(t) - 3.
\end{equation}
Here $k(t)=\langle v_\alpha^{4}\rangle/\langle v_\alpha^{2}\rangle^{2}$ is the kurtosis of a single Cartesian velocity component $v_\alpha$ (averaged over components and particles); the identity $\kappa=3\overline{A}^{2}$ and the two-bath result $\kappa=3c^{2}$ are quoted in this same one-dimensional, per-component convention. Lastly, to characterize the relaxation process, we compute the time averaged
excess kurtosis over a time interval. The latter is thus explicitly defined as
\begin{equation}
\kappa :=
\frac{1}{t_f - t_0}
\int_{t_0}^{t_f} \left[k(t)-3\right] dt.
\end{equation}
Finite-size fluctuations of the kurtosis in Gaussian samples have a variance scaling as
$\mathrm{Var}(k)\sim 24/N$. For $N\approx1500$ particles this implies
statistical fluctuations of the order of $\sim 0.1$. As we will detail, the excess kurtosis values that we find
when using stochastic thermostats are notably larger ($\kappa \sim 0.3$ for Two baths and $\kappa \sim 0.2$ for the HTLA discussed in the next section). These values exceed
pure sampling noise and represent genuine deviations from canonical
velocity statistics. This deviation underscores the viable  appearance of non-Maxwellian velocity distributions in supercooled systems.

\begin{figure}
    \centering
    \includegraphics[width=\columnwidth]{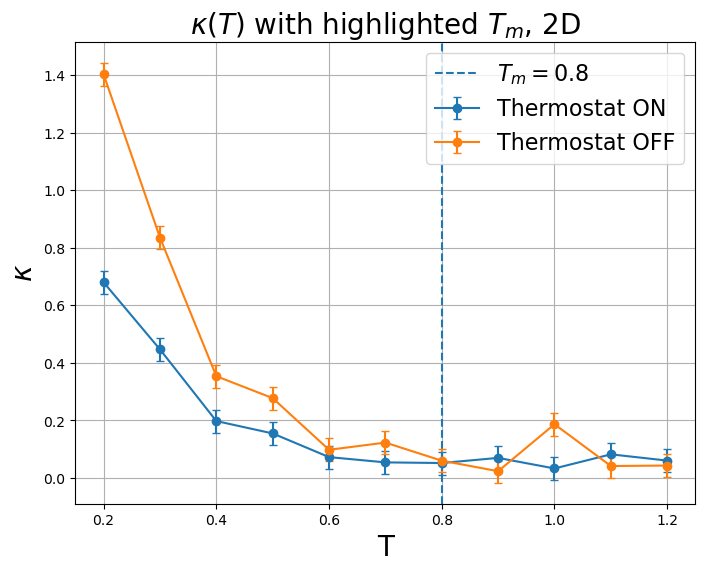}
    \caption[2D kurtosis]{ The excess kurtosis $\kappa$ of the temperature $T$ in a monodisperse two-dimensional Lennard-Jones system. $\kappa$ is measured after 10 MD time units from the end of the quenching scheme. The orange points are taken without a thermostat's participation in the relaxation of the system while blue ones are from measurements with a Langevin thermostat active throughout the entire process. It is evident that, when supercooled, the system is more outlier prone without the participation of the thermostat.}
    \label{fig:2Dkurt}
\end{figure}

The time series displayed in Fig.~\ref{fig:combined_kurtosis_structure} illustrate that, as is indeed known, conventional continuous thermostats reduce $\kappa$ to zero during equilibration. By contrast, stochastic thermostats {\it maintain a finite positive} $\kappa$ even after the temperature has stabilized. In this stabilized regime, the system effectively equilibrates inasmuch as all conventional metrics (including the intermediate scattering function) point to. Monodisperse two-dimensional Lennard-Jones systems offer a particularly instructive case: their temperature remains stable for approximately 100~MD time units without active thermostat coupling, allowing direct comparison of dynamics with and without thermostatic control. Because such a small two-dimensional sample can be cooled with a strongly coupled thermostat and then completely decoupled for the initial relaxation without compromising the stability of the temperature, we can isolate the mean kurtosis of the system during relaxation and how it is affected by the presence of the thermostat. The results indicate that the velocity distributions of the supercooled systems progressively deviate from being Gaussian as the temperature is lowered. This effect is much less apparent when the thermostat is kept on and points to the canonical constraint imposed by conventional thermostats actively masking the intrinsic non-equilibrium broadening. Following a removal of the thermostat, the system displays substantially elevated transient kurtosis, revealing that canonical constraints suppress intrinsic non-equilibrium fluctuations (Fig.~\ref{fig:2Dkurt}).

These observations establish $\kappa$ as a practical and
dimensionless measure of non-equilibrium velocity-space
broadening.

\subsection{Stochastic Non-Equilibrium Thermostats: HTLA and Two-Bath}

In addition to conventional equilibrium thermostats, we employed two
collision-based stochastic schemes designed to sustain controlled
non-equilibrium velocity sampling: a Heavy-Tailed variant of the Lowe--Andersen (HTLA)
thermostat \cite{Lowe,Andersen_thermostat} and the Two-Bath form of the Andersen thermostat \cite{Andersen_thermostat}.
Both the HTLA and the Two-Bath Andersen thermostat belong to the Andersen class of stochastic collision
thermostats, in which particle velocities are intermittently reassigned
through pairwise or single particle collision events occurring with a
specified rate. Unlike continuous drag based schemes, these thermostats
introduce strong stochastic perturbations. Consequently, these baths generate
broader than Maxwellian  instantaneous velocity distributions even when the mean kinetic
temperature is fixed.

The Two-Bath thermostat achieves non-equilibrium steady states by
randomly thermalizing particles to one of two heat reservoirs at
different temperatures. The HTLA thermostat generalizes the
Lowe--Andersen framework by drawing post collision velocities from
heavy-tailed distributions, thereby enhancing the probability of
high-velocity excursions and amplifying velocity heterogeneity.

In both cases, the system reaches a stationary state with constant mean
temperature but non-Maxwellian velocity statistics, making these
thermostats suitable for systematically tuning the degree of
non-equilibrium sampling. A detailed derivation of the algorithms, parameter choices, and
statistical properties of these thermostats is provided in Appendix \ref{app:thermostats}. 

\begin{figure*}[t]
    \centering
    \begin{subfigure}{0.48\textwidth}
        \centering
        \includegraphics[width=0.9\linewidth]{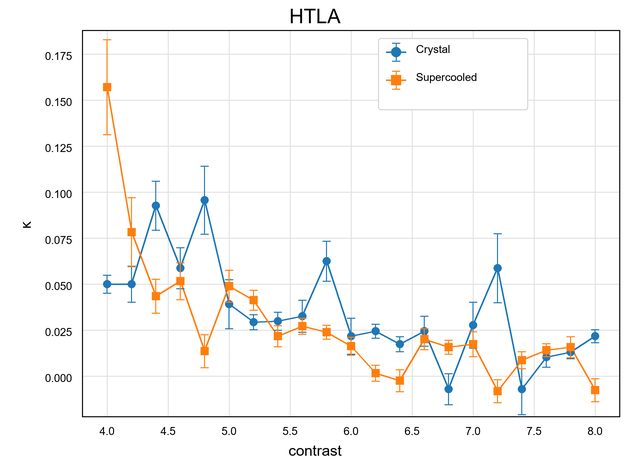}
        \caption{HTLA: $\kappa$ vs Contrast}
        \label{fig:kappa_htla_contrast}
    \end{subfigure}
    \hfill
    \begin{subfigure}{0.48\textwidth}
        \centering
        \includegraphics[width=0.9\linewidth]{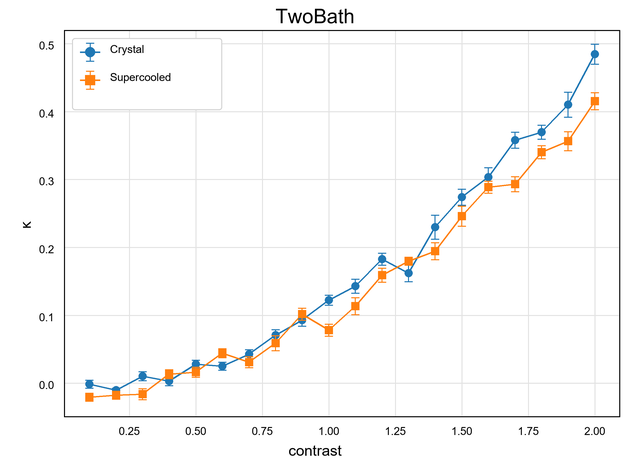}
        \caption{Two-Bath Andersen: $\kappa$ vs Contrast}
        \label{fig:Kappa_twobath_contrast}
    \end{subfigure}
    \caption[$\kappa$ vs contrast]{
    Dependence of velocity-distribution kurtosis $\kappa$ on thermostat contrast
    for two stochastic schemes in a 3D Lennard--Jones system for both supercooled (orange) and crystal (blue) samples.
    Left: Heavy-Tailed Lowe-Andersen (HTLA), where decreasing contrast enhances superstatistical broadening and produces systematic growth in excess kurtosis.
    Right: Two-Bath Andersen thermostat, where contrast induces bimodal thermalization and generates non-Gaussian velocity statistics whose kurtosis depends on the temperature separation and collision frequency.
    Both thermostats can modulate the statistical profile of the system, but the Two-Bath thermostat accesses a broader range of $\kappa$ values with less noise.
    }
    \label{fig:kappa_contrast_comparison}
\end{figure*}

The degree of non-equilibrium broadening accessible to each scheme is summarized in Fig.~\ref{fig:kappa_contrast_comparison}, which shows the dependence of the steady state excess kurtosis $\kappa$ on the thermostat contrast parameter for both schemes in a three-dimensional Lennard--Jones system, for supercooled (orange) and crystalline (blue) samples. For the HTLA thermostat (left panel), decreasing the contrast enhances the superstatistical broadening and produces a systematic growth of $\kappa$; for the Two-Bath thermostat (right panel), the temperature contrast induces bimodal thermalization whose kurtosis grows with the temperature separation and collision frequency. Both schemes modulate the statistical profile of the system, but the Two-Bath thermostat accesses a broader range of $\kappa$ values with less noise, which is why it is the primary tool used in the crystallization study below.

\begin{figure*}[t]
    \centering
    \begin{subfigure}{0.48\textwidth}
        \centering
        \includegraphics[width=\linewidth]{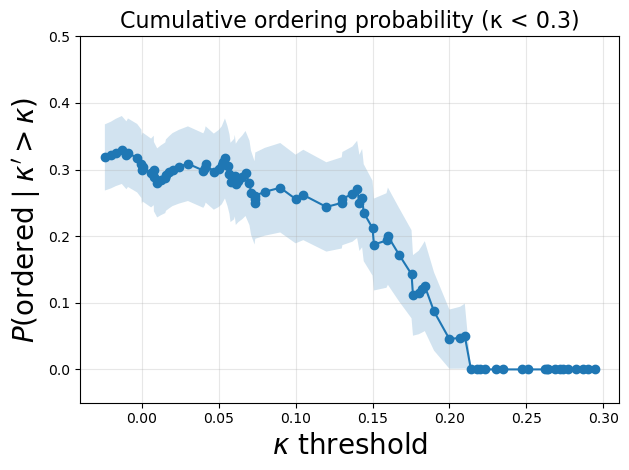}
        \caption{}
        \label{fig:cum_pcryst_80}
    \end{subfigure}
    \hfill
    \begin{subfigure}{0.48\textwidth}
        \centering
        \includegraphics[width=\linewidth]{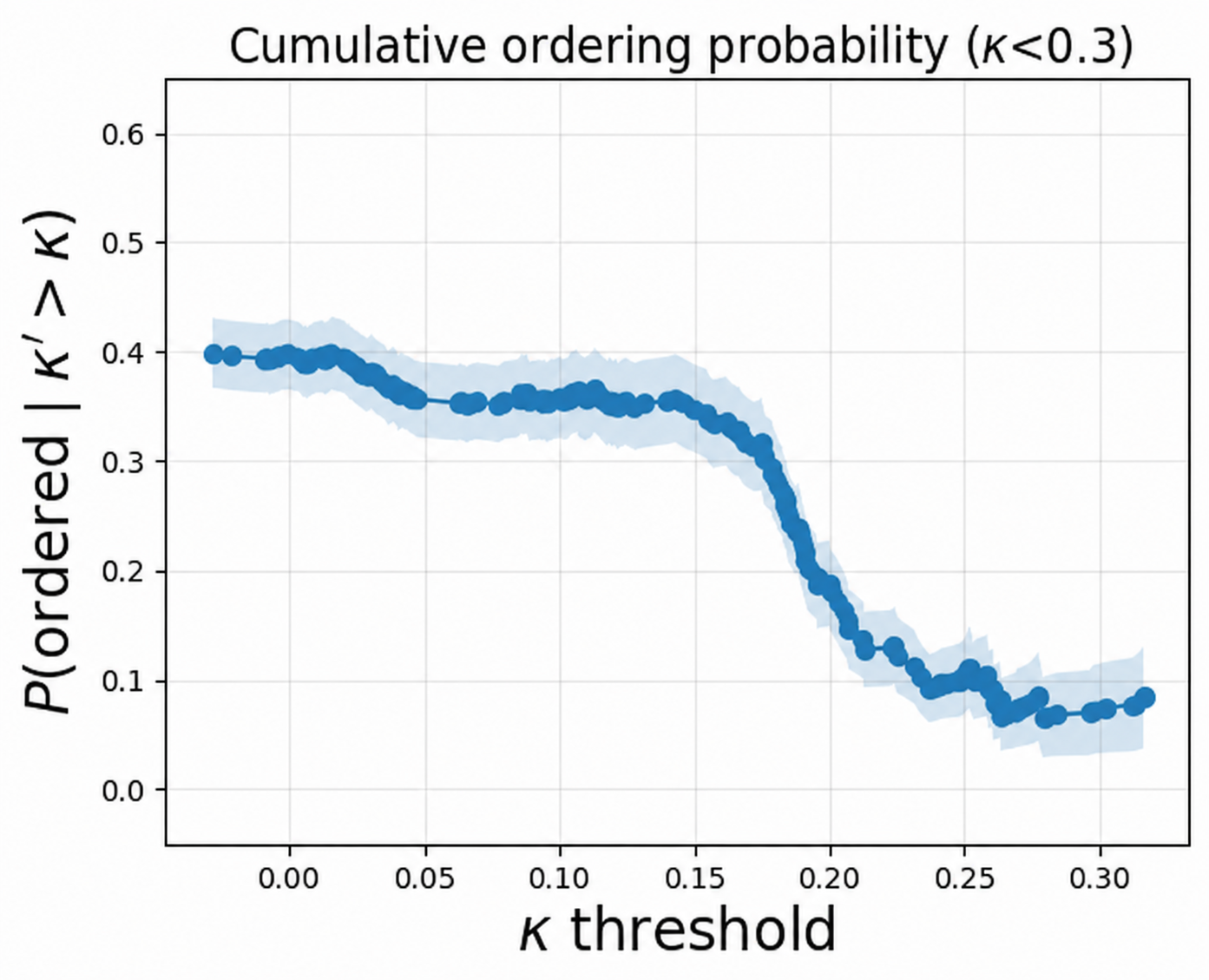}
        \caption{}
        \label{fig:cum_pcryst_120}
    \end{subfigure}
    \caption{
    The cumulative probability distribution for local bond orientational order conditioned on the velocity-kurtosis threshold.
    For each threshold $\kappa$, the curves show
    $P(\mathrm{ordered}\mid \kappa' \ge \kappa)$.
    Panel (a) the results for 80 runs; panel (b) the corresponding result for 150 runs.
    Shaded bands: binomial standard error.
    Increased kurtosis suppresses crystallization.
    Both panels are restricted to the accessible range $\kappa\lesssim0.3$ (the horizontal axis runs only to $\kappa\simeq0.3$ in each); the suppression of ordering therefore occurs entirely within the $\kappa\lesssim0.3$ window set by the stochastic thermostats.}
    \label{fig:cum_pcryst_combined}
\end{figure*}

\begin{figure}[t]
    \centering
    \includegraphics[width=1\linewidth]{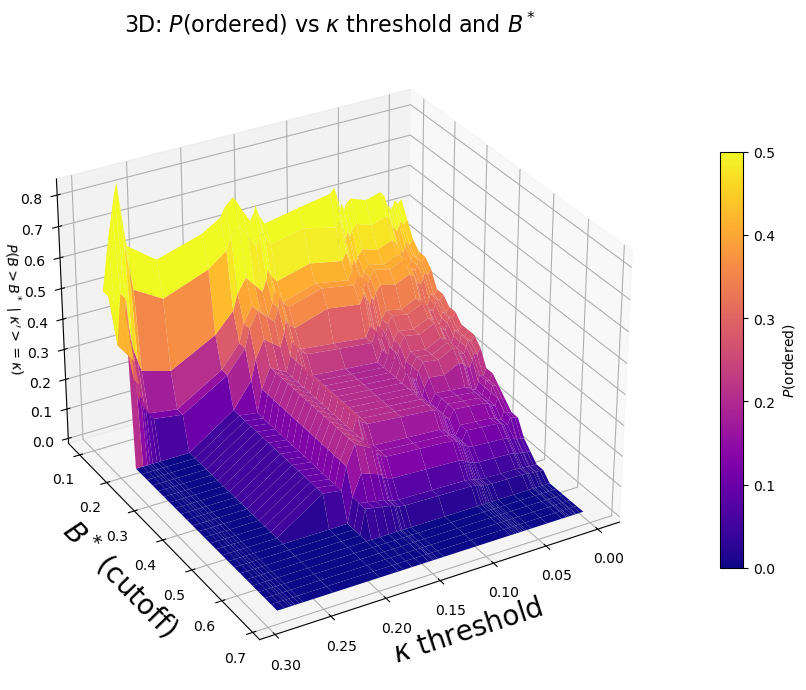}
    \caption[P(local Order) $B^*$ threshold]{
    Three-dimensional representation of the crystallization probability
    $P(\mathrm{cryst})$ as a function of excess velocity kurtosis $\kappa$
    and bond order threshold $B_{\mathrm{th}}$.
    The surface illustrates how increasing non-equilibrium velocity broadening
    (larger $\kappa$) systematically reduces the likelihood of crystallization
    across structural thresholds.
    Regions of high $B_{\mathrm{th}}$ emphasize stringent crystallinity criteria,
    while low thresholds capture more weakly ordered states.
    The data in the $\kappa =0$ plane coincide, identically,  with those found when using the conventional (i.e., non heavy tailed) Lowe-Andersen thermostat.
    }
    \label{fig:Pcryst_3D_surface}
\end{figure}

\subsection{Excess Kurtosis and Suppression of Crystallization}
\label{sec:2.4}

We now examine how the momentum or, equivalently, the velocity-space broadening affects structural
outcomes. Across $\sim$ 100 independent 3D supercooling runs for 3 different quenching schemes each. The number of particles $N=1500$, the time step $dt=0.004$, mass of the particles set  to $m=1$, the Boltzmann constant $k_{B}=1$, the density $\rho=1.1$, and the Lennard-Jones potential coefficients $\bar{b}=1$, $U=1$ all in natural MD units. The degree of polydispersity in the potential being $8\%$. The only control variable in our calculations was the strength of
non-equilibrium velocity sampling introduced by the thermostat.

Providing the average over a hundred runs, Figure~\ref{fig:cum_pcryst_combined} depicts the cumulative probability $P(\mathrm{ordered} \mid \kappa' \ge \kappa)$ that the cooled system develops a  sufficient degree of local order (i.e., that its bond orientational order parameter exceeds a minimal threshold value $B^*$) given that the excess kurtosis of the velocity was larger than $\kappa$. As is readily seen, when an excess kurtosis of $\kappa \sim 0.15$ is achieved, the probability of crystallization decreases sharply.
In fig.~\ref{fig:cum_pcryst_combined},
the lower cutoff on the local bond orientation order $B^* = 0.25$ was chosen so as to have maximum contrast between systems enjoying a higher degree of local order than those that do not.  Fig.~\ref{fig:Pcryst_3D_surface} shows the variation of the cumulative distribution for local bond orientational order over the full range of $B^*$ values that we examined. The use of threshold values of $B^*$ is not common practice in the literature; earlier investigations, e.g., ~\cite{Lechner2008,tenWolde1996} similarly sought to determine optimal threshold values on the local bond orientational order parameter and its generalization to best delineate between crystalline and supercooled states. All of the data points collected in Figs. \ref{fig:cum_pcryst_combined} and \ref{fig:Pcryst_3D_surface} are for supercooled fluid systems. For crystalline systems (inasmuch as crystallinity can be inferred in the simulated polydisperse systems), the local bond orientational order parameter was sizable (of values $B \gtrsim 0.7$ similar to those shown for the orange curve in Fig. \ref{fig:combined_kurtosis_structure} (b)). The trend that we find with increasing values of $\kappa$ indicates that broader velocity distributions actively suppress structural ordering.
Intermittent high velocity excursions disrupt coherent particle rearrangements required for nucleation,
maintaining configurational disorder even at identical mean temperatures.

Complementary spatial diagnostics reinforce this conclusion.
We contrast the coarse grained potential energy landscapes
for stochastic versus equilibrium thermostats.
Supercooled runs exhibit pronounced spatial heterogeneity,
while equilibrium crystallization produces uniform,
low-energy domains. The effect persists across system sizes ranging from 1500 to 10000 particles and is robust under periodic boundary conditions, indicating that it is not an artifact of finite size or rigid walls. We note in this connection that, when a rigid (rather than periodic) boundary condition is used, natural crystallization is predominantly nucleated at the walls of the system, producing an uneven radial distribution of the structural variables; this was the practical reason for adopting conventional periodic boundary conditions throughout. The wall-nucleation effect persists even when the system size is increased by an order of magnitude (the particle number was varied between $1500$ and $12000$). While this is a genuine finite-size effect caused by the rigid boundaries, it is also a clear illustration of the system exhibiting radically different structural outcomes for two different thermostatting protocols at the same nominal temperature. 
There is additional evidence suggesting the dependence of the
block-averaged structural relaxation time $\langle \tau_{\alpha} \rangle$
on the excess velocity kurtosis $\kappa$. We can see emergence of positive correlation: as $\kappa$ increases,
$\langle \tau_{\alpha} \rangle$ systematically grows.
In particular, beyond $\kappa \gtrsim 0.2$ the increase becomes sharper, indicating that progressively heavier-tailed
velocity statistics are accompanied by a marked slowing down of
$\alpha$-relaxation dynamics.
Since larger $\kappa$ quantifies stronger deviations from a Maxwellian
velocity distribution, this result demonstrates that non-Gaussian
fluctuations in the kinetic sector couple directly to long time
structural relaxation in the supercooled regime. A complementary discussion of the standard non-Gaussian parameter $\alpha_2(t)$, its relation to the $\alpha$-relaxation time $\tau_\alpha$, and the connection between dynamical heterogeneity and the crystallization time $\tau_{\mathrm{xtal}}$ is given in Appendix~\ref{app:nongaussian}.

\begin{eqnarray}
&& P(\mathrm{ordered} \mid \kappa' \ge \kappa)
:=
P\!\left( B \ge B^* \;\middle|\; \kappa' \ge \kappa \right) \nonumber
\\ && =
\frac{
\sum\limits_{r=1}^{N_{\mathrm{runs}}}
\Theta\!\left(B_r - B^*\right)
\Theta\!\left(\kappa'_r - \kappa\right)
}{
\sum\limits_{r=1}^{N_{\mathrm{runs}}}
\Theta\!\left(\kappa'_r - \kappa\right)
},
\end{eqnarray}
where $\Theta(\cdot)$ denotes the Heaviside step function that is equal to unity when its argument is positive and is zero for negative arguments. 

\section{ Determining Temperature Distributions From Single Particle Velocities}
\label{sec:ilt}
\label{cpt2}

The velocity distributions discussed in the last chapter allow for much more than merely observing that the steady state dynamics are non-Maxwellian. With their aid, we can infer the distribution of effective equilibrium temperatures that describe the nearly stationary supercooled state.

The velocity degrees of freedom do not fully equilibrate before structural arrest/crystallization in deeply supercooled systems. Figs. \ref{fig:cum_pcryst_120} and \ref{fig:Pcryst_3D_surface} demonstrate that excess kurtosis remains finite in the metastable supercooled state and does not relax to zero prior to crystallization (the underlying non-Gaussian dynamics, dynamical heterogeneity, and crystallization are reviewed in Appendix~\ref{app:nongaussian}). The thermostats that we employ emulate a supercooled system with self-consistent finite excess kurtosis. This establishes that non-Gaussian velocity statistics are an intrinsic and persistent feature of the supercooled state on physically relevant timescales.

A central result of a phase space formulation of the long lived nearly stationary state concerns a universality that allows for predictions. 

We reiterate and underscore that this universality is not an assumption but rather a consequence of the theorem of Ref. \cite{Nussinov2024} (Appendix \ref{app:theorem}). Specifically, in any system for which the canonical and microcanonical ensemble descriptions coincide, the reduced few body local density of an arbitrary stationary state (including, trivially, the equilibrium states themselves) is a weighted superposition of the respective reduced few body local equilibrium probability densities when the latter are summed over varying values of the state variables. Such ensemble equivalence is satisfied by a broad class of disorder-free Hamiltonians. This includes, notably, supercooled liquids and their crystalline and liquid counterparts~\cite{Nussinov2024}. Thus, for these systems, the few body correlations of \emph{any} stationary state are fully encoded by thermal equilibrium. In particular, within the stationary state, the long time local reduced $n$-body probability density (with the particle number $n$ being of order unity) at inverse temperature $\beta$, chemical potential $\mu$, and any other state variables is a weighted superposition of reduced probability densities of the equilibrium system \cite{Nussinov2024} at effective inverse temperatures $\beta'$, chemical potentials $\mu'$, etc. That is, 
\begin{equation}
\label{eq:localP}
\begin{split}
\rho_{{\sf stationary},n}(\beta, \mu, \ldots)
= \int d\beta'\, d\mu' \cdots \\
P(\beta, \mu, \ldots \mid \beta', \mu', \ldots)\;
\rho_{{\sf eq.},n}(\beta', \mu', \ldots)
\end{split}
\end{equation}

In general, an additional contribution arises from the phase-coexistence regime where the energy density and number density may vary continuously at fixed thermodynamic fields $\beta',\mu', \ldots$ \cite{Nussinov2024}; the full expression including this coexistence region is given in Eq.~(\ref{central-theorem}) of Appendix \ref{app:theorem}. The normalized weight $P$ acts as a \emph{spectral decomposition over equilibria}. This transforms the problem of characterizing complex non-equilibrium and metastable structures into one governed entirely by equilibrium statistical mechanics. Given the form of Eq. (\ref{eq:localP}) for the reduced local probability density, predictions can be made for numerous local observables if $P$ were known. We return to a central point underlying our work: the single particle velocity distributions allow for an extraction of the dependence of probability kernel $P$ on the effective inverse temperature $\beta'$ in a manner that is completely {\it independent} of the specific details of the system such as the potential energy describing the interactions between its constituents, details of chemical composition, and all structural facets. Theoretically, given the single particle velocity distribution over all velocities, one could infer $P(\beta|\beta')$ uniquely via a simple inverse Laplace transform. We turn to this aspect next.

\subsection{Theoretical Framework}

We briefly return to the theoretical framework of Ref. \cite{Nussinov2024} and focus on temperature fluctuations. As repeatedly emphasized above, in all classical equilibrium systems the velocities follow a Maxwellian distribution which is set solely by the temperature and no other equilibrium state variables. This implies that the probability distribution for Cartesian velocity components of a nearly stationary system at nominal inverse temperature \( \beta \) can be written as a convex hull average of Maxwellians at effective equilibrium inverse temperatures \( \beta' \) of the equilibrium system \cite{Nussinov2024},
\begin{equation}
P(|v_x|)
=
\sqrt{\frac{2m}{\pi}}
\int_0^\infty d\beta'\;
\beta'^{1/2}
\exp\!\left(- \frac{m \beta' v_x^2}{2}\right)
\, P(\beta \mid \beta').
\label{eq:pvx_mixture}
\end{equation}
Upon defining the Laplace transform variable
$s := \frac{m v_x^2}{2}$,
Eq.~\eqref{eq:pvx_mixture} immediately reads
\begin{eqnarray}
\label{eq:pvx_laplace}
&& P(\lvert v_x \rvert)
=
\sqrt{\frac{2m}{\pi}}
\int_0^\infty d\beta'\,
e^{-s\beta'}\ g(\beta'),\nonumber
\\ &&
g(\beta') := \beta'^{1/2} P(\beta \mid \beta').
\end{eqnarray}
Performing an inverse Laplace transform with respect to \(s\) therefore yields
\begin{equation}
P(\beta \mid \beta')
=
\sqrt{\frac{\pi}{2m}}\,
\beta'^{-1/2}\,
\mathcal{L}^{-1}_{s\to\beta'}
\!\left[
P(|v_x|)\Big|_{v_x=\sqrt{2s/m}}
\right].
\label{eq:inv_laplace_beta_prime}
\end{equation}
This formulation provides a direct method
to extract the effective inverse-temperature weighting kernel \(P(\beta \mid \beta')\) from the measured
velocity distribution \(P(|v_x|)\).
A practical implementation of this inversion and its comparison to measured \(P(v_x)\) is shown in Figs.~\ref{fig:Pbeta_combined} and \ref{fig:bessel_high_kurt}. Although the single particle distribution $P(\vec{v})$ is rotationally invariant, it is not a product of independent Cartesian-component distributions. This non-factorizability of the non-Maxwellian velocity distribution is deeply related to the constrained correlated dynamics in the supercooled liquid (Appendix \ref{app:correlated}).

\subsection{Numerical Inverse Laplace Transform via the Cohen Algorithm}

In practice, in order to extract $P(\beta \mid \beta')$, we will numerically invert the
Laplace transform relating $P(|v_x|)$ to the rescaled function
$g(\beta') = \beta'^{1/2} P(\beta \mid \beta')$.
Because numerical Laplace inversion is exponentially ill-conditioned, we employ the Cohen algorithm~\cite{Cohen2007}, which is based on a
Fourier--series representation of the Bromwich integral with controlled
convergence acceleration. In this method, the inverse transform is
approximated as
\begin{eqnarray}
&& g(\beta')
\approx
\frac{e^{A_{cohen}/2}}{2\beta'}
\sum_{k=0}^{N}
(-1)^k \, c_k \,
\Re\!\left[
\tilde{P}(s_k)
\right], \nonumber
\\ \quad
&& s_k = \frac{A_{cohen} + i\pi k}{2\beta'},
\end{eqnarray}
where $\tilde{P}(s)$ denotes the Laplace-domain function obtained from
$P(|v_x|)$ after removing the prefactor $\sqrt{2m/\pi}$ and substituting
$v_x = \sqrt{2s/m}$. The parameter $A_{cohen}>0$ controls numerical stability,
$N$ is the truncation order, and $c_k$ are Cohen acceleration coefficients
that suppress truncation error. The parameters $(A_{cohen},N)$ are chosen such
that convergence is achieved within the statistical uncertainty of the
measured velocity histogram.

In practice, we first smoothen  $P(|v_x|)$ using spline interpolation to
obtain a continuous representation and  then evaluate the transform at the
complex arguments required by the algorithm. The recovered function
$g(\beta')$ is subsequently rescaled according to
$P(\beta \mid \beta') = \beta'^{-1/2} g(\beta')$
and normalized to satisfy
$\int_0^{\infty} P(\beta \mid \beta') \, d\beta' = 1$.
Once $P(\beta|\beta')$ is known, we can determine numerous quantities for which the temperature is the prominent governing state variable of the supercooled fluid, including the viscosity as a function of temperature.

\subsection{Gamma-Distributed Effective Temperature}

A fundamental result of equilibrium statistical mechanics is that the maximization of the entropy subject to the condition of a given mean energy yields the canonical ensemble probability distribution with the inverse temperature serving as the Lagrange multiplier enforcing that constraint \cite{Jaynes1957}. The supercooled liquid
is not described by the canonical ensemble probability density with the system Hamiltonian. The canonical density built from the system Hamiltonian at a temperature below melting describes the equilibrium \emph{solid}, not the supercooled fluid, and the supercooled liquid simply does not have ``sufficient time'' to reach this global equilibrium state. Under rapid supercooling, different local regions of the system instead ``manage'' to equilibrate and maximize the entropy locally to varying temperatures, so that a vanishing variance of the effective temperatures $T'$ would correspond to the canonical density of the equilibrium solid rather than to the supercooled fluid. Instead, the supercooled liquid may only achieve a maximum entropy state
with a finite variance of the effective equilibrium temperatures $T'$. We thus seek the distribution $P(T|T')$ having a maximal Shannon entropy subject to: fixed average $\int_0^\infty T' P(T|T') dT' = T$; fixed finite standard deviation; support restricted to positive temperatures; and continuous vanishing at zero.

\begin{figure}[tbp]
    \centering
    \includegraphics[width=1\linewidth]{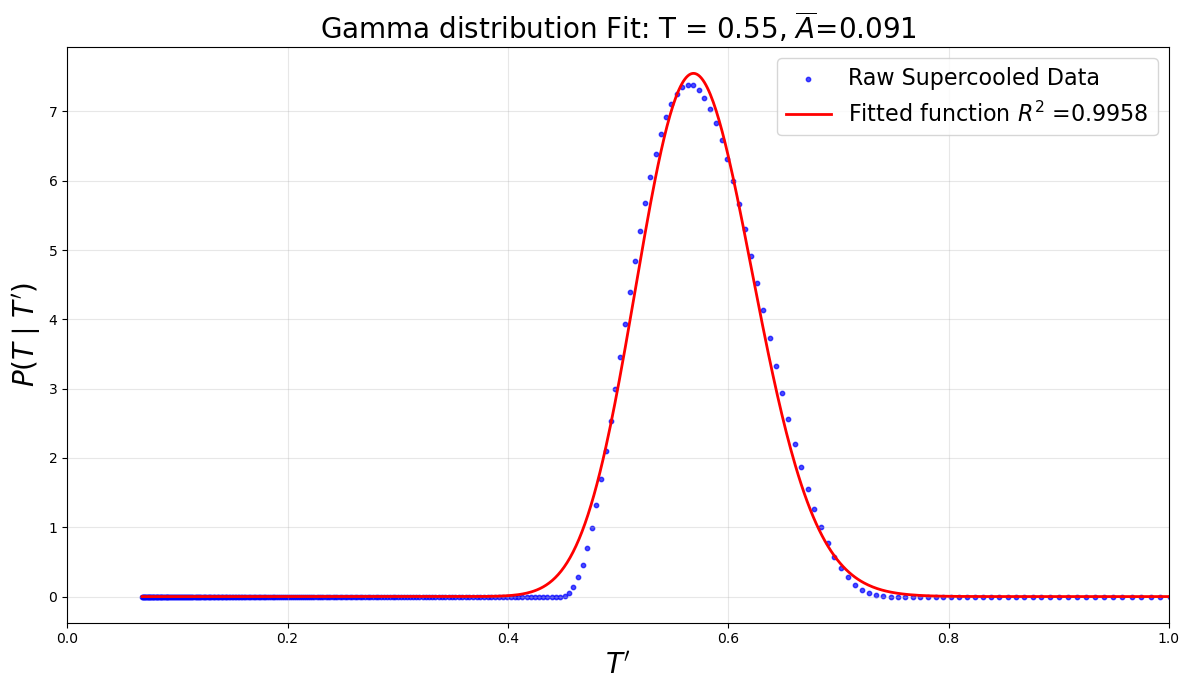}
    \caption[$P(T|T')$ Gamma fit]{ The Gamma distribution fit for the supercooled liquid temperature distribution \(P(T|T')\).}
    \label{fig:pvx}
\end{figure}

\begin{figure}[tbp]
\includegraphics[width=\linewidth]{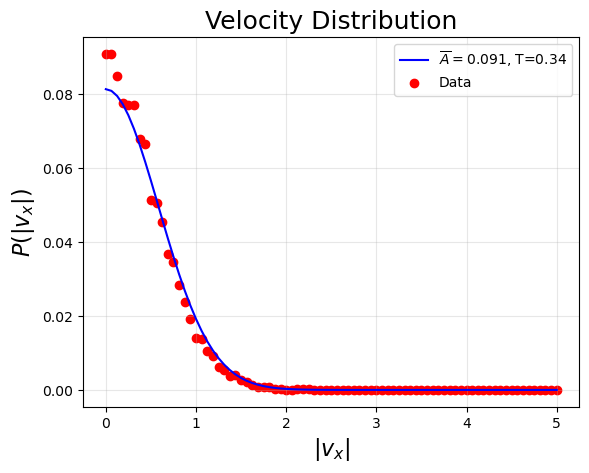}
        \caption[$P(|V_x|)$ comparison]{
            Comparison between the extracted velocity probability density $P(|v_x|)$ from raw data (red dots) to the distribution reconstructed from the Laplace transform of the Gamma distribution $P(T|T')$ of Eq. (\ref{eq:gamma_T}) (continuous blue curve).  The agreement demonstrates the quality of the gamma fit (Eq. (\ref{eq:PTv_gamma_closed})). For the computation of the viscosity, the nontrivial high velocity (effective high  $T'$) tail is of greater importance than the peak $v_x=0$ value. The sample shown here is at mean temperature $T=0.34$, which lies below the glass transition temperature $T_g\simeq0.4$. Our effective Gamma distribution $P(T|T')$ fit forms indeed appear to extend to low temperatures. 
        }
        \label{fig:bessel_high_kurt}
\end{figure}

Maximizing the Shannon entropy
${\sf H} = - \frac{1}{\ln 2} \int_0^\infty P(T|T') \ln P(T|T') \, dT'$
subject to the above constraints yields a Gamma distribution, 
\begin{equation}
P(T|T')
=
\frac{(T{'})^{{\sf A}-1}}{\Gamma({\sf A})\,{\sf B}^{{\sf A}}}\,
\exp\!\left(-\frac{T'}{\sf{B}}\right),
\qquad T'>0 .
\label{eq:gamma_T}
\end{equation}
The two lowest cumulants of the distribution are
$\langle T^{'}\rangle = {\sf A B}$,
$\mathrm{Var}(T^{'})={\sf A B}^{2}$,
and the ratio of the standard deviation to the mean is a constant,
\begin{equation}
\frac{\sigma_{T^{'}}}{\langle T^{'}\rangle}=\frac{1}{\sqrt{{\sf{A}}}} := \overline{A}.
\label{eq:gamma_moments}
\end{equation}
Thus ${\sf B}$ sets the temperature scale, while ${\sf A}$ controls the relative width of the distribution and quantifies the degree of thermal heterogeneity. In the limit ${\sf{A}}\to\infty$ at fixed mean temperature, the distribution collapses to
$P(T|T')\to\delta(T-T')$, corresponding to the equilibrium system.
The quality of the Gamma fit is demonstrated in Fig.~\ref{fig:pvx}. Taken together, Figs.~\ref{fig:pvx} and \ref{fig:bessel_high_kurt} show that the Gamma form captures the high-energy (large $T'$) tail of $P(T|T')$ (which dominates the non-equilibrium, fluid-like contributions to averages such as the viscosity) more faithfully than the low-$T'$ (slower particle) region.

\paragraph{Inverse temperature distribution.}
Transforming to the inverse temperature $\beta=1/T$ yields
\begin{equation}
P(\beta|\beta')
=
\frac{\beta'^{({\sf{A}}+1)}}{\Gamma({\sf{A}})\,{\sf{B}}^{{\sf{A}}}}~
e^{-1/{({\sf{B}}\beta')}},
\label{eq:inv_gamma_beta}
\end{equation}
an inverse--Gamma distribution in $\beta'$.

\subsection{Velocity Distribution with Gamma Temperature Prior}

When $P(T|T')$ is chosen to be a Gamma distribution,
\begin{equation}
P_T(v_{ix})
=
\int_{0}^{\infty}
\sqrt{\frac{m_i}{2\pi T'}}\,
\exp\!\left(-\frac{m_i v_{ix}^2}{2 T'}\right)\,
P(T|T')\, dT' .
\label{eq:PTv_gamma_def}
\end{equation}
Substituting Eq.~\eqref{eq:gamma_T} and employing
$s = m_i v_{ix}^2/(2)$, the integral can be evaluated analytically using
standard Bessel function identities:
\begin{equation}
P_T(v_{ix})
=
\sqrt{\frac{m_i}{2\pi}}\;
\frac{2}{\Gamma({\sf A})\,{\sf B}^{\sf A}}\,
(s {\sf B})^{\frac{{\sf A}-\frac{1}{2}}{2}}\,
K_{{\sf A}-\frac{1}{2}}\!\left(2\sqrt{\frac{s}{{\sf B}}}\right).
\label{eq:PTv_gamma_closed}
\end{equation}
Here, $K_\nu(z)$ is the modified Bessel function of the second kind.
In the equilibrium limit ${\sf A}\to\infty$ with ${\sf AB}=T$, the mixture collapses to
the Maxwell--Boltzmann distribution. For small $\overline{A} \ll 1$, the Gamma distribution becomes a Gaussian,
\begin{eqnarray}
\label{approxGauss}
P(T|T') \approx \frac{1}{T \overline{A}\sqrt{2\pi} } \exp\left[-\frac{1}{2} \left(\frac{T' - T}{T \overline{A}} \right)^2 \right].
\end{eqnarray}
Beyond the maximum entropy motivation \cite{Jaynes1957} for the Gamma distribution form of $P(T|T')$, there is a further practical benefit to this choice over other distributions that behave similarly near their peak (for example a log-normal distribution in $T'$): unlike a log-normal prior, the Gamma prior leads to the closed-form expression of Eq.~\eqref{eq:PTv_gamma_closed}, which is particularly convenient for analytical analysis and parameter fitting.

\begin{figure}[tbp]
    \centering
    \includegraphics[width=1\linewidth]{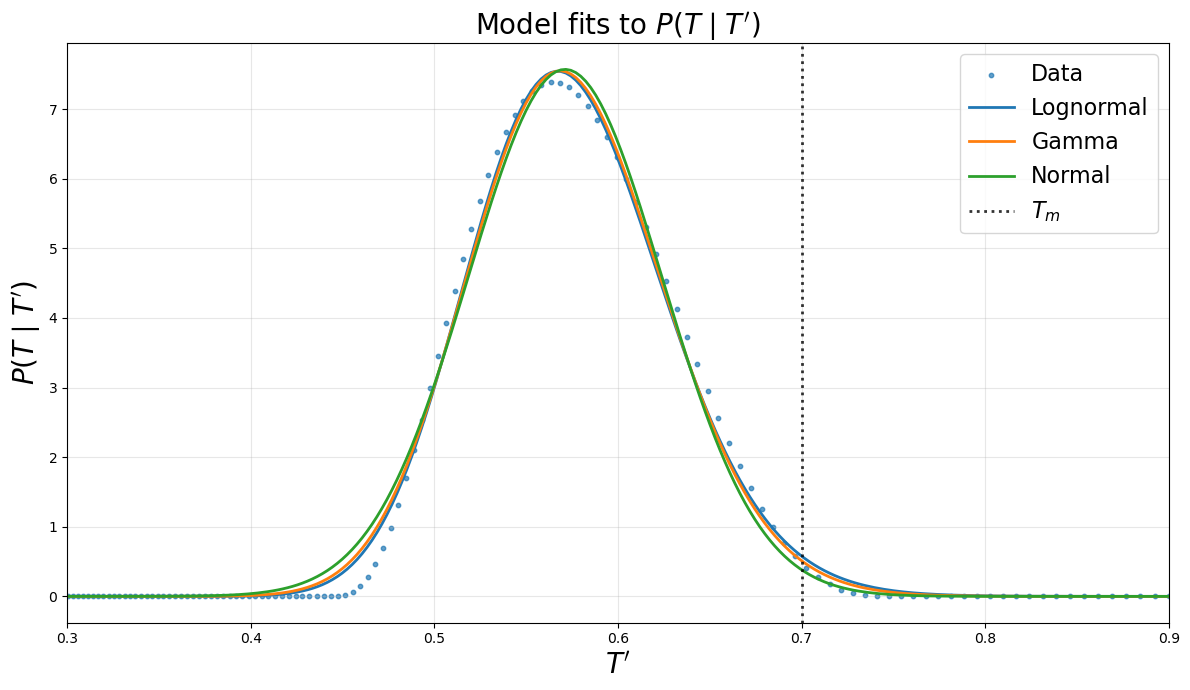}
    \caption[$P(T|T')$ 3 fits comparison]{Tail only fits for $T'>T_m=0.7$. Curves show least-squares fits of Gamma, Normal, and Lognormal models to $P(T\mid T')$ restricted to the high-$T'$ region. Legend labels report the RMSE values computed on the tail window: Gamma (0.0413), Normal (0.0494), Lognormal (0.0573).}
    \label{fig:tail_fits_rmse}
\end{figure}

The discriminating physical information, however, lies in the high-$T'$ tail rather than at the peak, since it is the weight at $T'>T_m$ that controls the equilibrium-fluid-like (e.g.\ viscous) contributions to averages performed with Eq.~(\ref{eq:localP}). To probe this directly we refit the candidate families \{Normal, Lognormal, Gamma\} using only the tail region $T'>T_m=0.7$. In this restricted domain the Gamma fit yields the smallest residual (root-mean-square error, RMSE), indicating that it most accurately captures the melting-side tail shape, although the differences between the functional forms are not large. 
The normal and log-normal fits feature standard deviations $\sigma_{T'}$ similar to that of the Gamma distribution, $\sigma_{T'}\approx\overline{A}\,T$. That is, the ratio of the standard deviation to the mean of the effective temperatures is the same fixed, temperature  independent dimensionless constant $\overline{A}$ while the long high $T'$ tail of $P(T|T')$ emulates that of an inverse-Gamma distribution. This tail only comparison is shown in Fig.~\ref{fig:tail_fits_rmse}.

\subsection{Relating $\overline{A}$ to Velocity Kurtosis}

Given Eqs. (\ref{eq:pvx_mixture},\ref{eq:inv_gamma_beta}),
\begin{equation}
    \langle v_x^2 \rangle
    =  \frac{1}{m} \left\langle\frac{1}{\beta^{'}} \right\rangle,~~
    \langle v_x^4 \rangle
    = \frac{3}{m^2} \left\langle \frac{1}{(\beta^{'})^2} \right\rangle.
    \end{equation}
Because the velocity distribution at each effective temperature is Maxwellian, $\langle v_x^4\rangle_{T'}=3\langle v_x^2\rangle_{T'}^2$; averaging over $P(T|T')$ then gives the kurtosis in closed form,
\begin{equation}
\label{ksim}
    k :=  \frac{\langle v_x^4\rangle}{\langle v_x^2\rangle^2}= 3\,\frac{\langle T'^2\rangle}{\langle T'\rangle^2}= 3\left(1+\overline{A}^2\right),
\end{equation}
so that the excess kurtosis $\kappa:=k-3$ is given by 
Eq. (\ref{eq:kappa_Abar}). Only the first two moments of $P(T|T')$ enter, so Eq.~(\ref{eq:kappa_Abar}) is exact and independent of the shape of the distribution; it fixes the relative variance of the effective-temperature distribution directly from the measured kurtosis, with no reference to the microscopic interactions.
Both the direct fitting of $P(\beta|\beta')$ and using Eq.~(\ref{ksim}) to approximately compute $\overline{A}$ from the kurtosis yield values that are very close to each other (up to relative corrections of 0.1 or raw changes in the values of $\overline{A}$ that are $\sim 0.01$). Furthermore, the error bars in computing $\kappa$ are themselves $\sim 0.01$. These error bars are many standard deviations below the finite $\kappa$ values that we find. Thus, our measured deviations from Maxwellian velocity statistics are not an artifact of  statistical noise.

\subsection{Statistical Interpretation of the Tail Area and its Connection to Relaxation Slowdown}

\begin{figure}[tbp]
    \centering
    \includegraphics[width=1\linewidth]{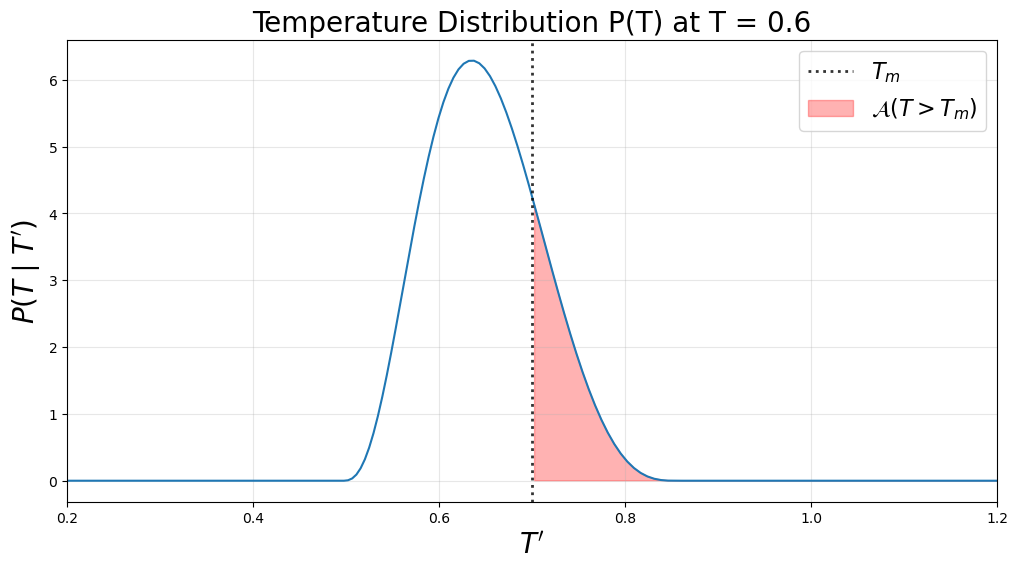}
    \caption[Tail Area Intro.]{
    Probability distribution $P(T|T')$ for the supercooled liquid at a  mean temperature $T=0.6$, fitted with a Gamma distribution.
    The vertical dashed line indicates the melting temperature $T_m$, and the shaded region corresponds to the high temperature tail $T^\prime>T_m$.
    The area of this region quantifies the excess non-equilibrium population above the melting threshold.}
    \label{fig:PT_tail_T06}
\end{figure}

At temperatures $T^{'} > T_{m}$, the local reduced few body probability density $\rho_{{\sf eq.},n}(\beta', \ldots)$ in Eq. (\ref{eq:localP}) is that of an equilibrium fluid above its melting temperature. Within the equilibrium liquid,
long time hydrodynamic flow may occur. By contrast, in the equilibrium solid, such dynamics are not possible \cite{sausset}. Thus, we define the
\emph{high temperature tail area} $\mathcal{A}(T>T_m)$ as the integrated probability mass of the
effective temperature distribution above the melting point:
\begin{equation}
\mathcal{A}(T'>T_m) := \int_{T_m}^{\infty} P(T|T')\, dT'.
\label{eq:tail_area_def}
\end{equation}
Figure~\ref{fig:PT_tail_T06} illustrates this construction for a representative supercooled liquid at mean temperature $T=0.6$. The vertical dashed line marks the melting temperature $T_m$, and the shaded region above it is the high temperature tail whose area quantifies the excess non-equilibrium population above the melting threshold.
We consider the
\emph{inverse tail area}, $\mathcal{A}^{-1}(T'>T_m)$, as a statistical proxy for the shear viscosity,
\begin{equation}
\eta(T) \; \simeq \; \frac{\eta_{\sf eq.}(T_{m}^{+})}{\mathcal{A}(T'>T_m)}.
\label{eq:eta_tail_relation}
\end{equation}
Equation~(\ref{eq:eta_tail_relation}) is obtained from the Stokes relation. In Refs.~\cite{grant16,grant15,preprint6,preprint2,Nussinov2024} the viscosity is read from the terminal velocity of an infinitesimal sphere released into the liquid under a uniform gravitational field, which by Stokes' law is inversely proportional to $\eta$. Only regions whose effective temperature places them in the equilibrium fluid  ($T'>T_{m})$ contribute to this flow. In an equilibrium solid the terminal velocity vanishes: a solid below its liquidus does not generate a drag that grows with velocity~\cite{sausset}, and although defects such as diffusing vacancies and gliding dislocations permit activated, plastic relaxation above a yield stress, they do not sustain continuous viscous flow under an infinitesimal stress. Where static crystalline domains coexist with fluid ones, the solid fraction forms a load-bearing skeleton with a finite yield stress, so infinitesimal stresses again drive no steady flow and the sphere does not fall. Averaging the terminal velocity over the effective-temperature distribution of Eq.~(\ref{eq:localP}) therefore retains only the weight above the liquidus, the high temperature tail area $\mathcal{A}(T'>T_m)$m and yields Eq.~(\ref{eq:eta_tail_relation}), the equilibrium liquidus being the sole surviving temperature scale (Appendix \ref{app:theorem}). 

\begin{figure*}[!t]
    \centering
    \begin{subfigure}[t]{0.49\textwidth}
        \centering
        \includegraphics[width=\textwidth]{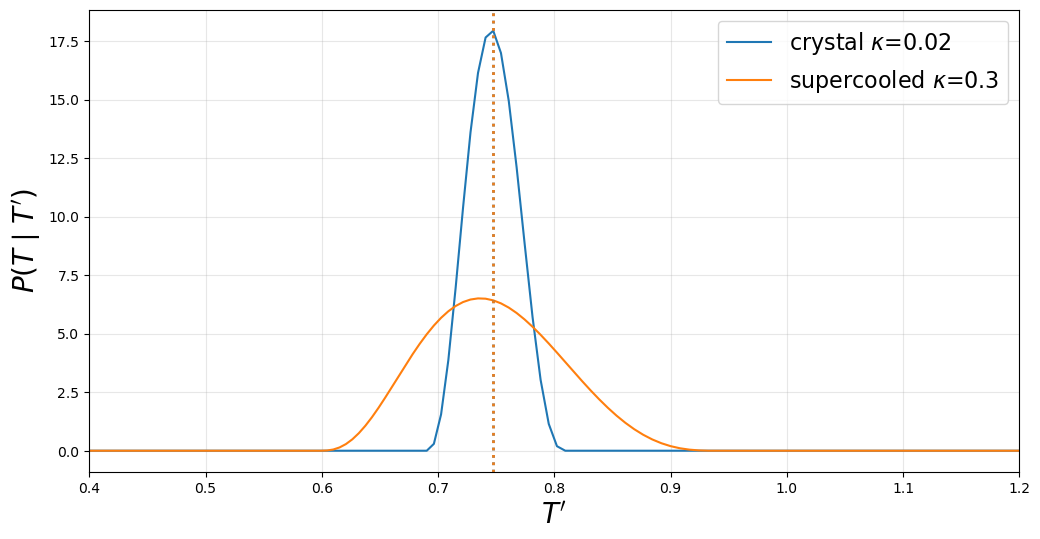}
        \label{fig:Pbeta_comparison}
    \end{subfigure}
    \hfill
    \begin{subfigure}[t]{0.49\textwidth}
        \centering
        \includegraphics[width=\textwidth]{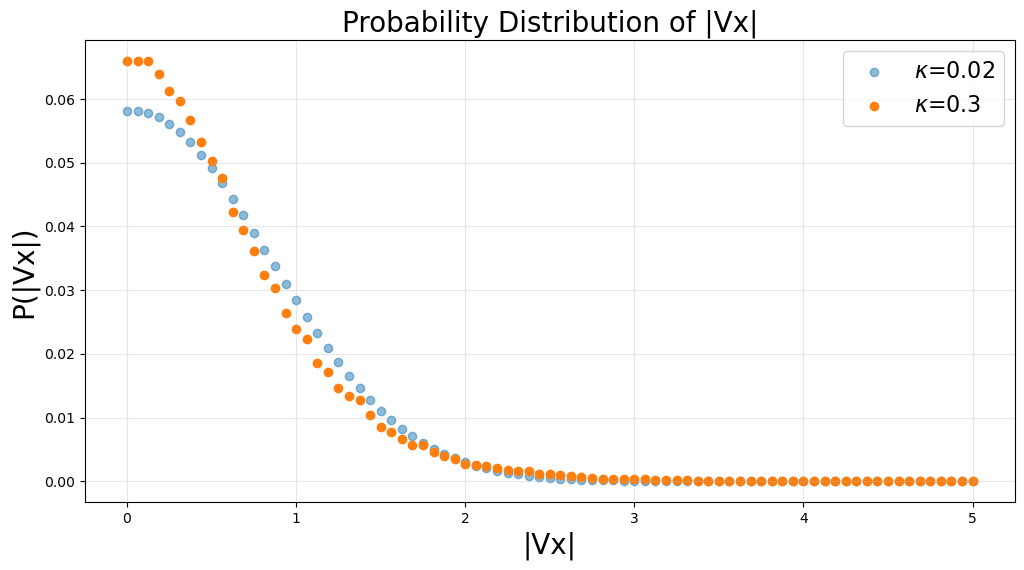}
        \label{fig:Pbeta_lognorm}
    \end{subfigure}
    \caption[Inverse Laplace for two $P(v_x)$ distributions]{
    (a) Probability distribution \(P(T|T')\) obtained via inverse Laplace transformation of \(P(v_x)\) using the Cohen algorithm, comparing equilibrium (blue) and supercooled (yellow) states.
    (b) Corresponding Normalized probability distributions \( P(|v_x|) \) vs \( |v_x| \) from an MD simulation of a supercooled liquid with Cumulative Kurtosis values $\kappa = 0.02,0.3$. Blue curve is from a supercooled liquid while the yellow stands for an equilibrium crystal.}
    \label{fig:Pbeta_combined}
\end{figure*}

\begin{figure*}[!t]
    \centering
    \includegraphics[width=\linewidth]{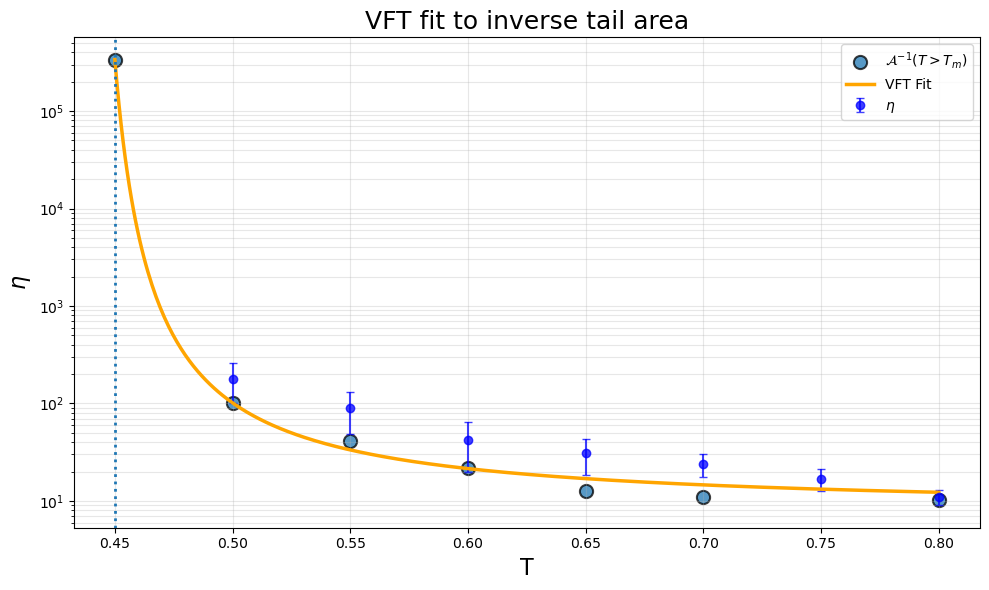}
    \caption[VFT]{
    Comparison between the reciprocal of the high temperature tail area of Eq. (\ref{eq:eta_tail_relation}) extracted from the temperature distributions \(P(T|T')\), its respective VFT fit approximation (Eq. (\ref{eq:VFT_law})), and the numerically measured viscosity. As the probability weight above the melting temperature decreases, the reciprocal \(1/\mathcal{A}\) increases in tandem with the rapid growth of viscosity captured by the phenomenological VFT law. The plotted viscosity points plotted were obtained, via the SLLOD algorithm, from the $\kappa=0.3$ ($\overline{A}\approx0.316$) supercooled sample whose temperature distributions and tail areas appear in Figs.~\ref{fig:Pbeta_combined} and \ref{fig:PT_and_tail_area}.}
    \label{fig:InvTailArea_VFTfit}
\end{figure*}

\begin{figure*}[!t]
    \centering
    \begin{subfigure}[t]{0.49\textwidth}
        \centering
        \includegraphics[width=\textwidth]{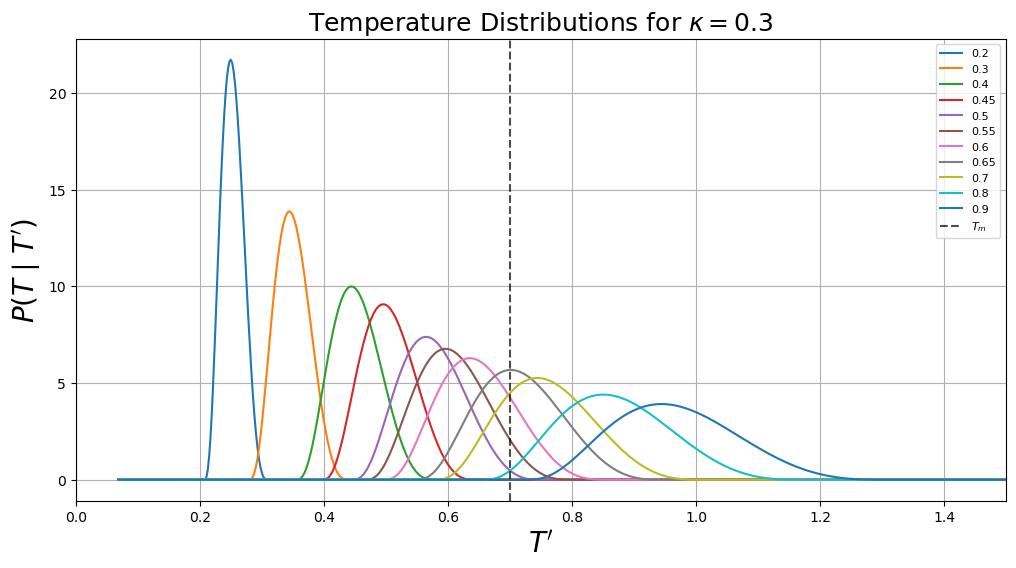}
        \label{fig:PT_distributions}
    \end{subfigure}
    \hfill
    \begin{subfigure}[t]{0.49\textwidth}
        \centering
        \includegraphics[width=\textwidth]{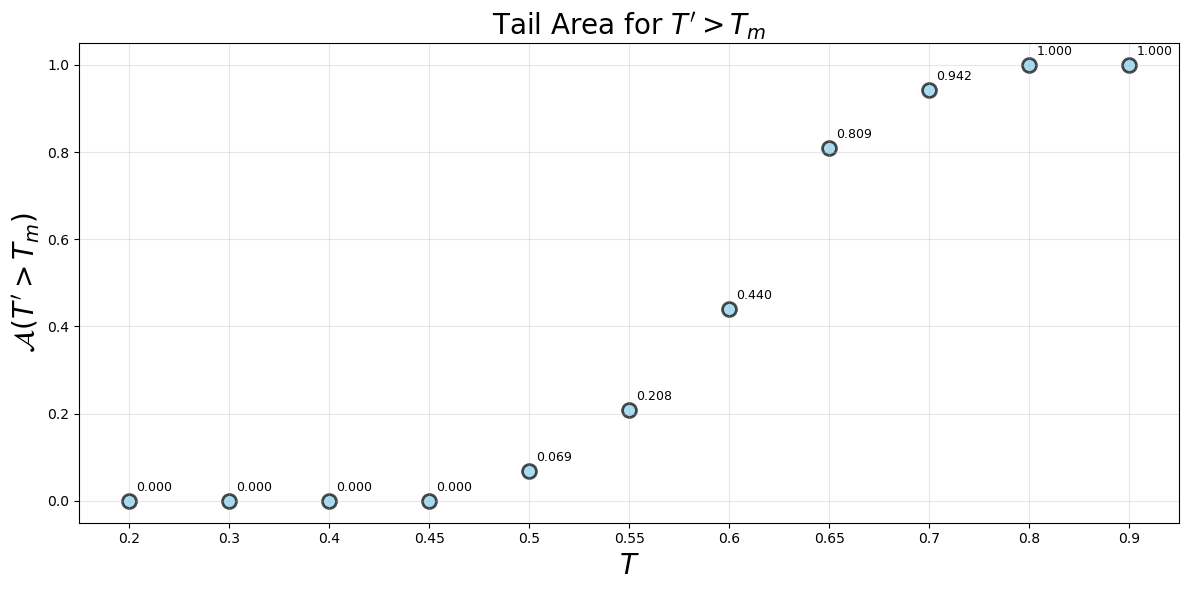}
        \label{fig:PT_tail_area}
    \end{subfigure}
    \caption[$\mathcal{A}(T^{'}>T_{m})$ vs Temperature]{ Left - distributions $P(T|T^{'})$ with $T_m$ highlighted by a dotted line. Right - Corresponding area above
    melting, $\mathcal{A}(T^{'}>T_{m})$, quantifying the fraction of probability mass of being above the equilibrium  melting temperatures for each case.
    }
\label{fig:PT_and_tail_area}
\end{figure*}

As the system temperature $T$ is supercooled below $T_m$, the weight associated with effective temperatures $T'$ that are larger than $T_{m}$ decreases sharply. This rapid decrease reflects the dominance of dynamically arrested regions as the temperature is lowered. The rapid growth of relaxation times with decreasing temperature makes a direct determination of the viscosity at low temperatures computationally prohibitive~\cite{HansenMcDonald}. We employed the VFT fit to extrapolate the inverse tail area to low temperatures. The computed viscosity was obtained directly from non-equilibrium MD using the SLLOD algorithm (Appendix \ref{app:sllod}). Fig. \ref{fig:InvTailArea_VFTfit} illustrates that the rise of the computed viscosity with decreasing temperature is approximately consistent with the VFT fit \cite{bib:vft,bib:vft2,bib:vft3} of Eq. (\ref{eq:VFT_law}). The above results constitute a consistency test of Eq. (\ref{eq:eta_tail_relation}). The viscosity data points in Fig.~\ref{fig:InvTailArea_VFTfit} and the temperature distributions and tail areas in Figs.~\ref{fig:Pbeta_combined} and \ref{fig:PT_and_tail_area} were all computed for the same supercooled polydisperse Kob-Andersen Lennard--Jones liquid of excess kurtosis $\kappa=0.3$  (i.e.,\ an effective width $\overline{A}=\sqrt{\kappa/3}\approx0.316$). We chose this larger $\kappa=0.3$ value specifically because its tail-area curve is more sizable than for smaller $\kappa$ and thus easier to track. The comparison between the kurtosis derived width and the independently measured SLLOD viscosity introduces no adjustable parameters tying the two together. 

As noted in Eq. (\ref{approxGauss}), for $\overline{A} \ll 1$, Eq.~(\ref{eq:eta_tail_relation}) may be approximated by
\begin{eqnarray}
\label{eq:visc1}
\eta(T) \simeq  \frac{\eta(T^+_{\sf melt})}{{\sf erfc}  (\frac{T_{\sf melt} - T}{\overline{A}T \sqrt{2}})},
\end{eqnarray}
with ${\sf erfc}(z)$ denoting the complementary error function. 

\begin{figure*}
	\centering
	\includegraphics[width=1.9 \columnwidth, height=1.3 \textheight, keepaspectratio]{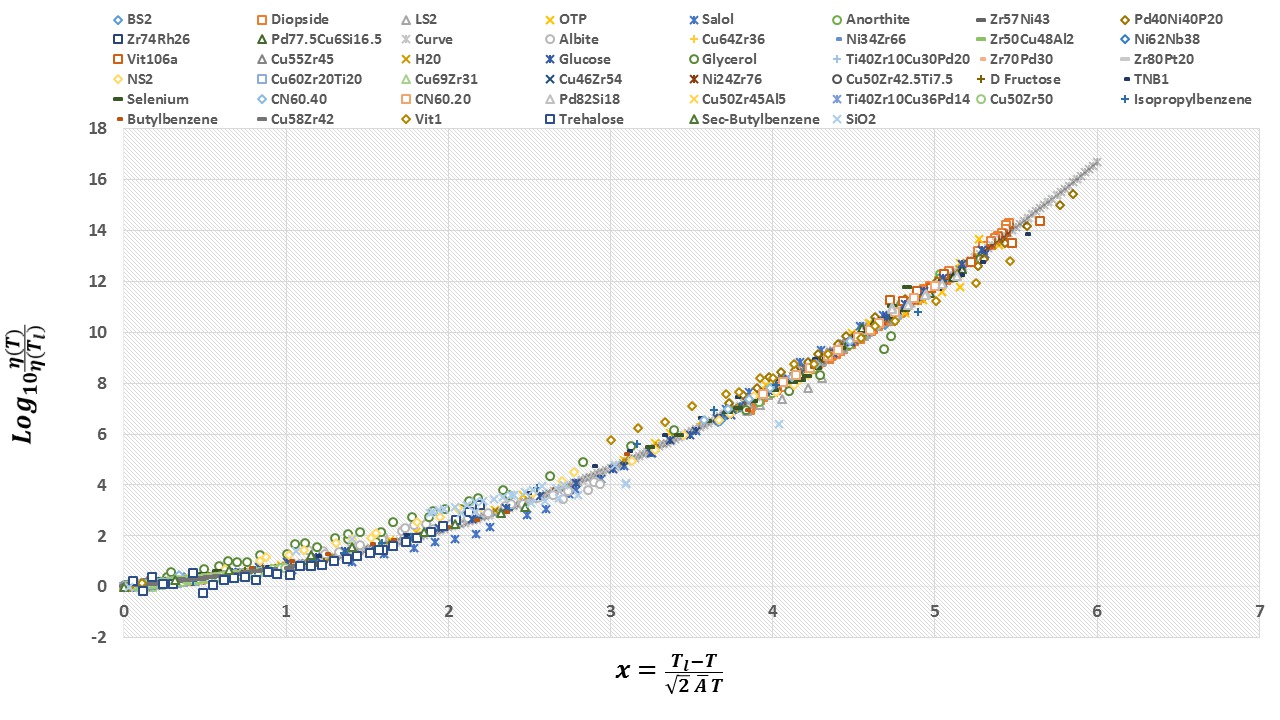}
	\caption[Viscosity]{
Reproduced from \cite{grant15}. An empirical test of Eq.~(\ref{eq:visc1}) across 45 supercooled liquids. Viscosities are scaled by the viscosity $\eta(T_{l})$ at the liquidus temperature  and plotted against the dimensionless variable $x := \frac{T_{l} - T}{\sqrt{2} \overline{A} T}$. Data spanning silicate, metallic, and organic systems --- covering the full range of kinetic fragilities --- collapse onto a single universal curve over 16 decades, suggesting a universality among glassforming liquids. No conjectured temperatures (such as the VFT ideal glass transition fitting  parameter $T_0$) appear. Rather the only temperature scale appearing in this collapse is the  experimentally accessible, equilibrium  liquidus temperature $T_{l}$.} 	\label{Collapse.}
\end{figure*}

We stress that in the above simulation comparison in which $\overline{A} \sim 0.3$, the results plotted in Fig.~\ref{fig:InvTailArea_VFTfit}) were computed with the \emph{exact} tail-area relation Eq.~(\ref{eq:eta_tail_relation}), the integrated high-$T'$ probability mass itself, rather than the $overline{A} \ll 1$ approximation of Eq.~(\ref{eq:visc1}). In fitting to experiments (Fig. \ref{Collapse.}), one typically finds $\overline{A} \ll 1$ \cite{grant15,preprint2}. For the few strongly broadened glass formers, e.g.,\ $\overline{A}\simeq0.234$ for Zr$_{57}$Ni$_{43}$, the exact relation Eq.~(\ref{eq:eta_tail_relation}) is the form to use although in the latter case, the fit was still performed with Eq. (\ref{eq:visc1}). Averaging over all of fitted fluids in Fig. (\ref{Collapse.}), the mean liquid dependent dimensionless parameter $\overline{A}$ is $0.08$ with a standard deivation of $0.03$. We underscore that these fits contain \emph{two} material parameters: the experimentally measured high temperature liquidus (melting) scale $T_l\approx T_m$ entering Eq.~(\ref{eq:visc1}) and the fitted dimensionless width $\overline{A}$. Both are required to reproduce the measured viscosities across all tested liquids, including those with varying degrees of compositional mixing among their constituents. 

Empirically, the same width $\overline{A}$ appears to be consistent with more than the temperature dependence of the viscosity alone. Because the distribution of the effective local temperatures $T^{'}$ has standard deviation $\sigma_{T_g}\simeq\overline{A}\,T_g$ at the glass transition, the hysteresis of the fixed pressure specific heat measured on heating and cooling through $T_g$ is predicted to span a window $\Delta T_h\approx2\sigma_{T_g}\approx2\overline{A}\,T_g$~\cite{Nussinov2024,preprint6}. For the four glass formers whose viscosity fits supply $\overline{A}$ (OTP, glycerol, salol, and TNB), published calorimetric data give the dimensionless ratio $\Delta T_h/(2\overline{A}\,T_g)\approx0.92\pm0.08$, qualitatively consistent with the predicted value of unity~\cite{Nussinov2024,preprint6}. As is seen in the 45 liquid collapse of Fig. \ref{Collapse.}~\cite{grant15,preprint2}, there are a few outliers that do not conform well to Eq. (\ref{eq:visc1}) Glycerol is a notable outlier in this collapse of and its specific heat likewise departs from that of other supercooled liquids; the latter feature has been suggested as evidence for a ``liquid--liquid'' transition~\cite{AngellheatE}. Other departures from the collapse include Vit1 and trehalose and SiO$_2$, an archetypal extremely ``strong'' glass former (in Angell's classification \cite{bib:angell}). Our simulations for the model polydisperse Kob-Andersen type system yield values of $\overline{A}$ consistent in magntiude to those found in experiment ~\cite{grant15,preprint2}. In our simulations the accessible non-equilibrium steady states extend from $\kappa=0$ up to a stability ceiling at excess kurtosis $\kappa\approx0.3$ (equivalently a width $\overline{A}=\sqrt{\kappa/3}\approx0.316$); larger values are associated with an inability to achieve a stable steady state).

The stochastic thermostats give access to an entire \emph{spectrum} of steady state kurtosis values rather than a single one. For the Two-Bath thermostat, this spectrum is bounded above by $\kappa\simeq 0.3$: below this value the system reaches a well-defined non-equilibrium steady state (and the stationarity assumptions underlying our analysis hold), whereas above $\kappa\simeq0.3$ those stationarity assumptions begin to break down and a stable steady state can no longer be maintained. The upper bound value of $\kappa\simeq0.3$ constitutes a \emph{stability ceiling}, i.e., it is the largest non-Gaussianity that can be stably sustained. We emphasize that this upper bound is \emph{not} a value to which the steady state is pinned. The supercooled system supports a stable steady state at \emph{any} kurtosis below this threshold- the thermostats can realize the whole range $0<\kappa\lesssim0.3$. We further find values of $\overline{A}$ within this range which are similar in magnitude to those in the data collapse of Fig. (\ref{Collapse.}).

The right panel of Fig.~\ref{fig:PT_and_tail_area} shows a sigmoid like dependence of the excess tail area \(\mathcal{A}(T^{'} > T_{m})\) on temperature. This smooth crossover spans the interval between the conventional melting temperature \(T_m \simeq 0.8\) and the glass transition temperature \(T_g \simeq 0.4\) for Lennard--Jones type systems. Importantly, the measured excess tail area arises exclusively from the non-equilibrium velocity distributions generated by the thermostat. In strict thermal equilibrium, velocity distributions are exactly Maxwellian, implying \(\mathcal{A}(T^{'} > T_{m})=0\) at all temperatures. In that limit, the sigmoid like curve collapses to a step function at \(T_m\), thereby eliminating any sensitivity to \(T_g\) as a meaningful crossover scale.

Concurrent with the enhanced velocity fluctuations relative to those in the equilibrium system (with these broader fluctuations associated with a spread of temperatures $T'$), other fluctuations are present. Crystallization in supercooled systems may proceed \cite{nucleation1,nucleation2} via correlated large scale fluctuations in which particles forming an entire shell simultaneously join a crystalline seed. 

\chapter{The Methods Used}

\subsection{Simulation Parameters}
We performed MD supercooling simulations using a $6$--$12$ Lennard--Jones
interaction, while varying the thermostatting protocol employed in the system. To suppress
crystallization, a slight polydispersity was introduced in the Lennard--Jones length scale.
The pairwise interaction potential between particles $i$ and $j$ separated by a distance
$r_{ij}$ is given by
\begin{equation}
  U_{ij}(r_{ij}) =
  4 U
  \left[
  \left(\frac{b_{ij}}{r_{ij}}\right)^{12}
  -
  \left(\frac{b_{ij}}{r_{ij}}\right)^{6}
  \right],
\end{equation}
where $U$ sets the interaction energy scale and $b_{ij}$ is an effective interaction length
defined from particle-specific size parameters $b_i$ and $b_j$ via the arithmetic
mixing rule $b_{ij} = \tfrac{1}{2}(b_i + b_j)$. The parameters $b_i$ are drawn from a Gaussian distribution centered around a mean value $\bar{b}$, introducing a controlled level of
polydispersity ($8$--$10\%$) while preserving the overall Lennard--Jones form of the interaction.

The relevant base parameters for which most of our simulations were conducted are: the number of particles $N= 500$--$10000$, the time step $dt=0.004$, mass of the particles $m=1$, the Boltzmann constant $k_{B}=1$, the density $\rho=1.1$, and the Lennard-Jones potential parameters: $U=1$ and mean $\bar{b}=1$ over the polydisperse system.

\begin{figure*}[tbp]
    \centering
    \begin{subfigure}{.5\textwidth}
        \centering
        \includegraphics[width=.8\linewidth]{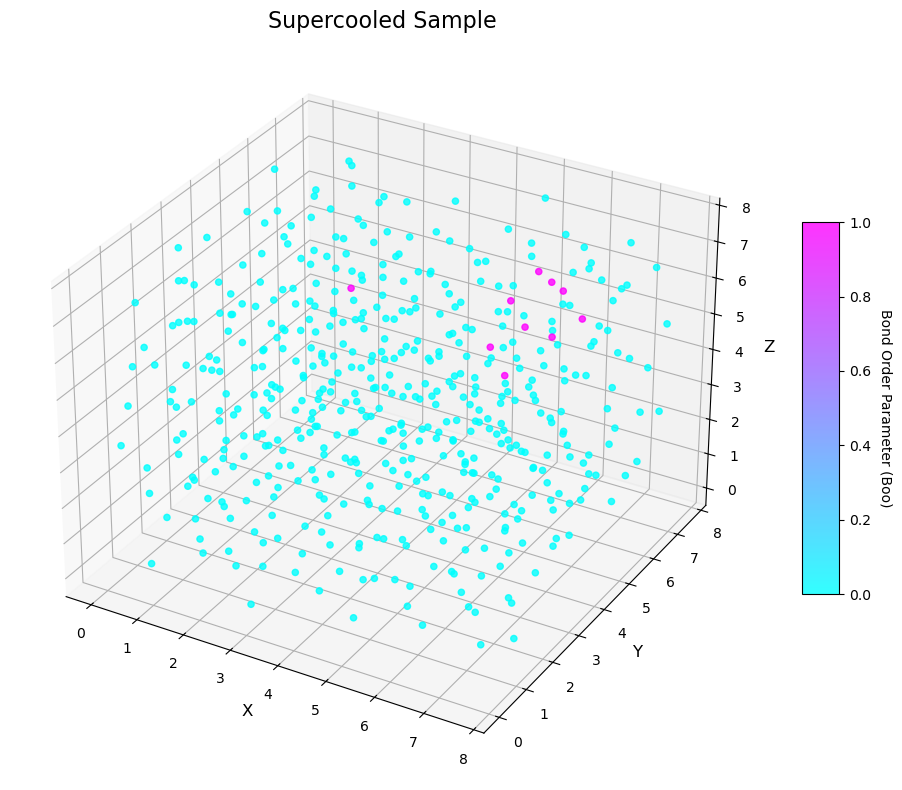}
        \caption{Bond order for the supercooled liquid}
        \label{fig:boo_supercooled}
    \end{subfigure}%
    \begin{subfigure}{.5\textwidth}
        \centering
        \includegraphics[width=.8\linewidth]{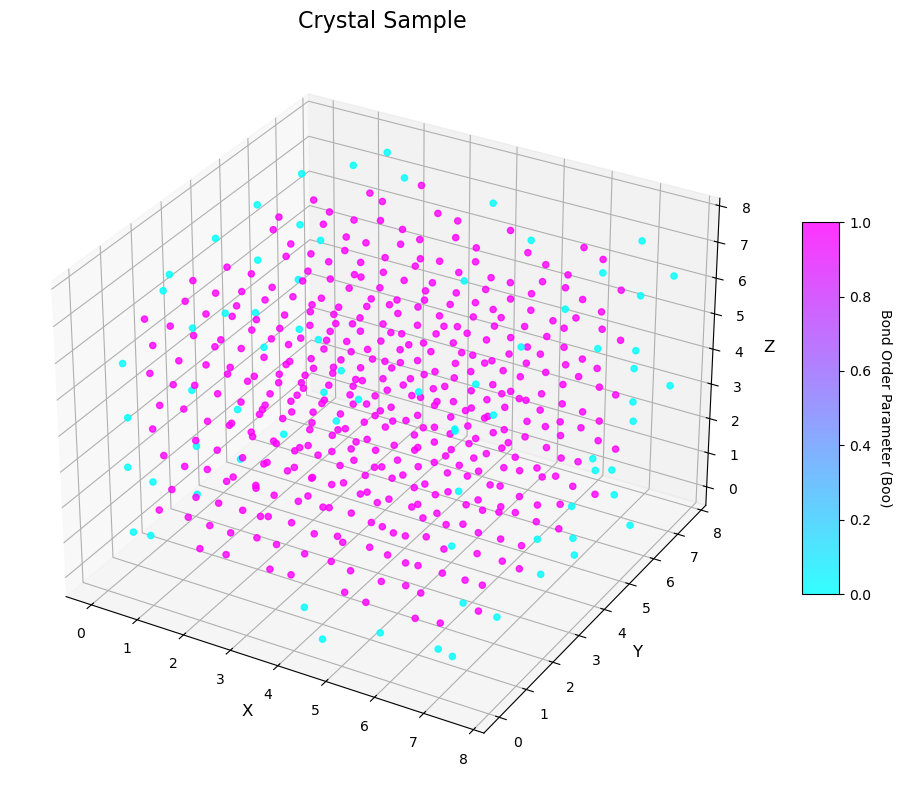}
        \caption{Bond order for the crystal}
        \label{fig:boo_crystal}
    \end{subfigure}
    \vspace{0.5em}
    \begin{subfigure}{.5\textwidth}
        \centering
        \includegraphics[width=.8\linewidth]{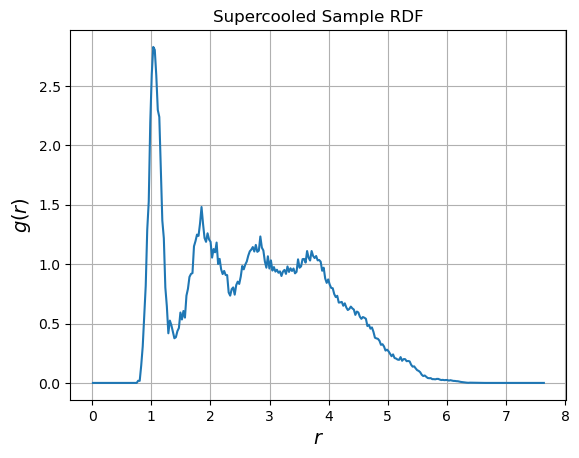}
        \caption{$g(r)$ for the supercooled liquid}
        \label{fig:rdf_supercooled}
    \end{subfigure}%
    \begin{subfigure}{.5\textwidth}
        \centering
        \includegraphics[width=.8\linewidth]{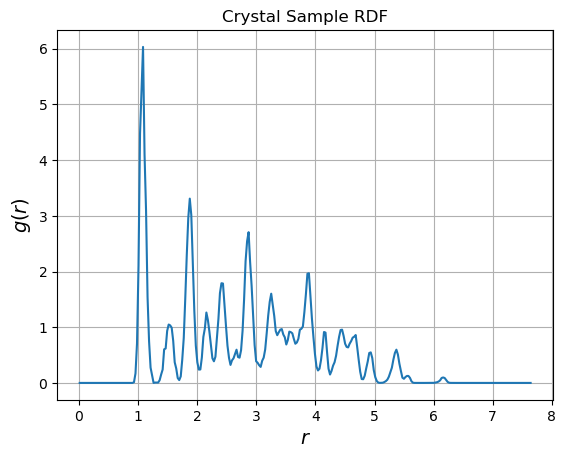}
        \caption{$g(r)$ for the crystal}
        \label{fig:rdf_crystal}
    \end{subfigure}
    \caption[Structure comparison]{
    Spatial distribution of local structural order and pair correlations for MD particles under different thermostatting conditions.
    \textbf{Top:} bond orientational order illustrating suppressed crystallinity in the supercooled liquid (low collision-rate thermostat) and long range order in the crystal (Berendsen thermostat).
    \textbf{Bottom:} Corresponding radial distribution functions $g(r)$, showing broadened and damped coordination shells in the supercooled state and sharp, long range oscillations characteristic of crystalline order.
    }
    \label{fig:Boo}
\end{figure*}

Figure ~\ref{fig:Boo} demonstrates two contrasting states achieved from this system at $T=0.5$ with $8\%$ polydispersity. It is integral for our study that the system is a near perfect crystal former ($B \approx 0.8$) while also being able to reach supercooled states of negligible structure under specific thermostats and quenching schemes. The equations of motion were integrated using the standard Velocity Verlet
algorithm, which is symplectic and time-reversible to second order in
the time step $dt$. At each step the new particle positions are first determined from the current velocities and forces, after which the forces are recomputed and the velocities updated accordingly; this scheme ensures good energy stability even in the absence of a thermostat and provides a consistent framework for incorporating the various thermostatting protocols considered here.

\subsection{Structure, Intermediate Scattering and Local Potential Energy}

The structure parameter is calculated as the six-fold bond orientation order parameter as codified in the celebrated work by Steinhardt, Nelson, and Ronchetti \cite{BOO}.
For each particle $i$, the complex bond-order coefficients are defined as
\begin{equation}
q_{lm}(i) =
\frac{1}{N_b(i)}
\sum_{j \in \mathcal{N}(i)}
Y_{lm}\!\left(\hat{r}_{ij}\right),
\end{equation}
where $Y_{lm}$ are spherical harmonics of degree $l$,
$\hat{r}_{ij}$ is the normalized bond vector,
$\mathcal{N}(i)$ denotes the neighbor set, and $N_b(i)$ is the number of
neighbors of particle $i$.

We focus on $l = 6$, which is particularly sensitive to crystalline
(FCC/HCP) symmetry. The rotationally invariant scalar order parameter
for each particle is
\begin{equation}
q_6(i) =
\left[
\frac{4\pi}{2l+1}
\sum_{m=-l}^{l}
\left| q_{lm}(i) \right|^2
\right]^{1/2}.
\end{equation}

The instantaneous average structural order over all particles is defined as
\begin{equation}
B(t) =
\frac{1}{N}
\sum_{i=1}^{N}
q_6(i,t).
\end{equation}

The self part of the intermediate scattering function $F(\vec{q},t)$ is calculated as:
\begin{equation}
F(\vec{q},t) := \frac{1}{N} \sum_{i=1}^{N} e^{i\vec{q} \cdot (\vec{r}_i(t)-\vec{r}_j(0))}.\end{equation}

\begin{figure}[tbp]
    \centering
    \includegraphics[width=1\linewidth]{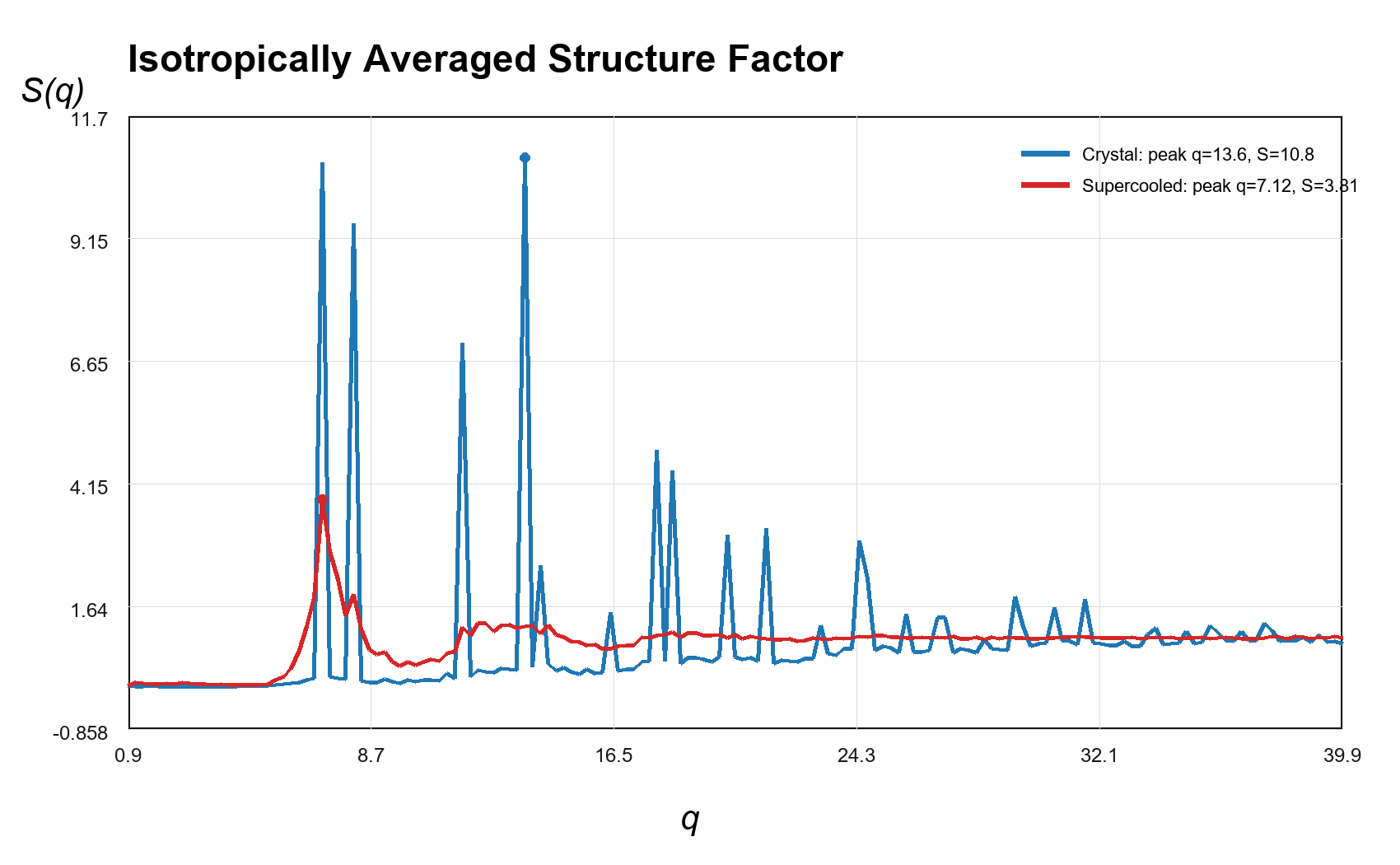}
    \caption[S(q)]{The overlaid structure factors $S(q)$ of the supercooled liquid and of the crystalline system.}
    \label{fig:Structure_Factor}
\end{figure}

Here, $N$ is the number of MD particles, $\vec{q}$ is the wave vector at which the intermediate scattering function is evaluated, $\vec{r}_i$ is the location of the $i$-th particle and $t$ is the time. The magnitude of the pertinent wave number $q$ at which $F$ is evaluated is the one maximizing the structure factor $S(|\vec{q}|=q)$ as seen for a crystal sample (Fig.~\ref{fig:Structure_Factor}). Measurements of  $F(q,t)$ evaluated at the wavenumber maximizing $S(q)$ are customarily employed to diagnose the stability and relaxation of the system~\cite{Kob-Andersen}.

\begin{figure*}[!t]
    \centering
    \begin{subfigure}{.48\textwidth}
        \centering
        \includegraphics[width=.9\linewidth]{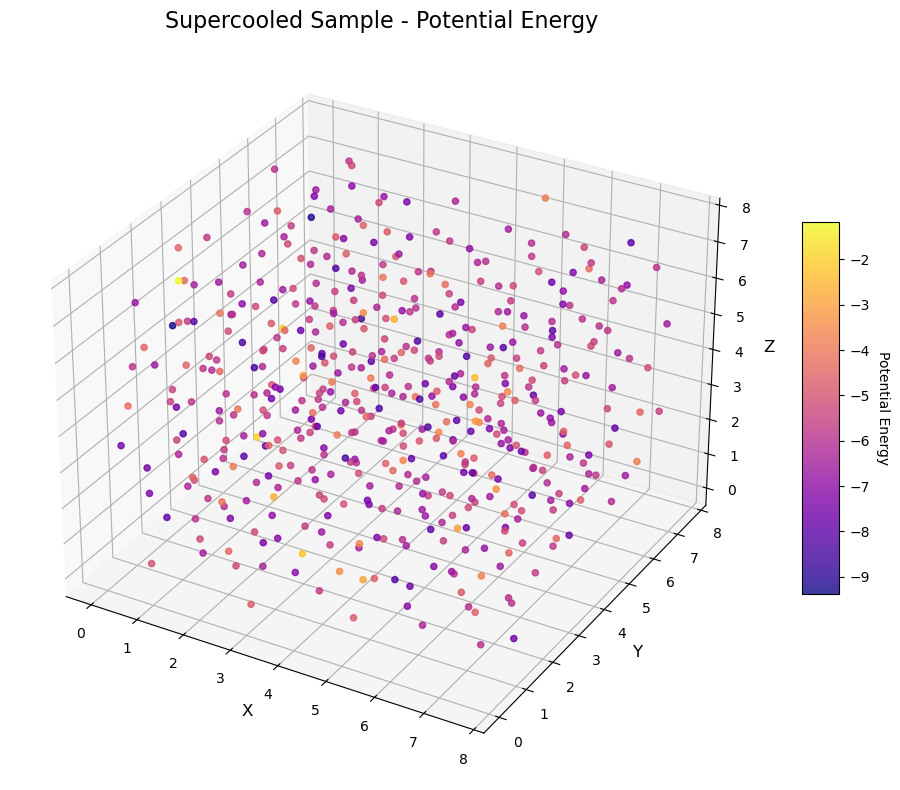}
        \caption{Spatial map (supercooled)}
        \label{fig:spatial_stoch}
    \end{subfigure}
    \hfill
    \begin{subfigure}{.48\textwidth}
        \centering
        \includegraphics[width=.9\linewidth]{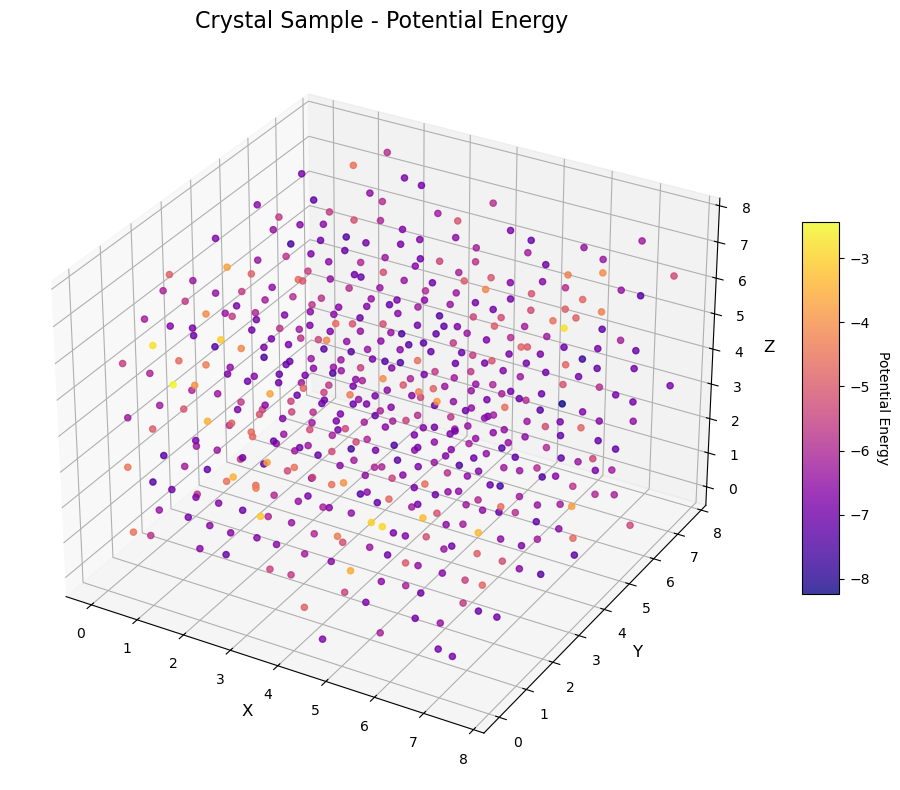}
        \caption{Spatial map (crystal)}
        \label{fig:spatial_lang}
    \end{subfigure}
    \vspace{0.6em}
    \begin{subfigure}{.48\textwidth}
        \centering
        \includegraphics[width=.9\linewidth]{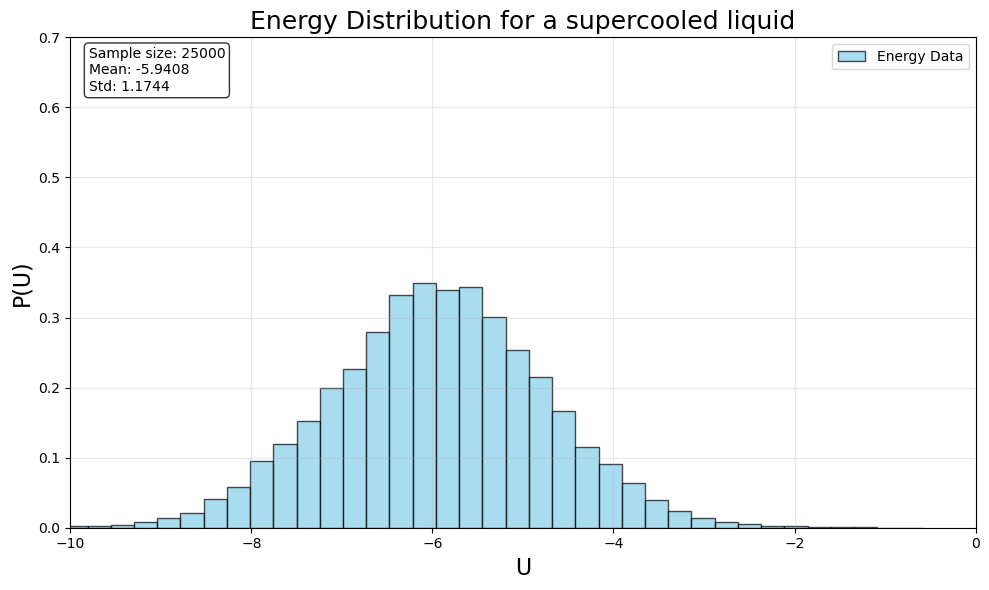}
        \caption{Per-particle $U$ distribution (supercooled)}
        \label{fig:hist_stoch}
    \end{subfigure}
    \hfill
    \begin{subfigure}{.48\textwidth}
        \centering
        \includegraphics[width=.9\linewidth]{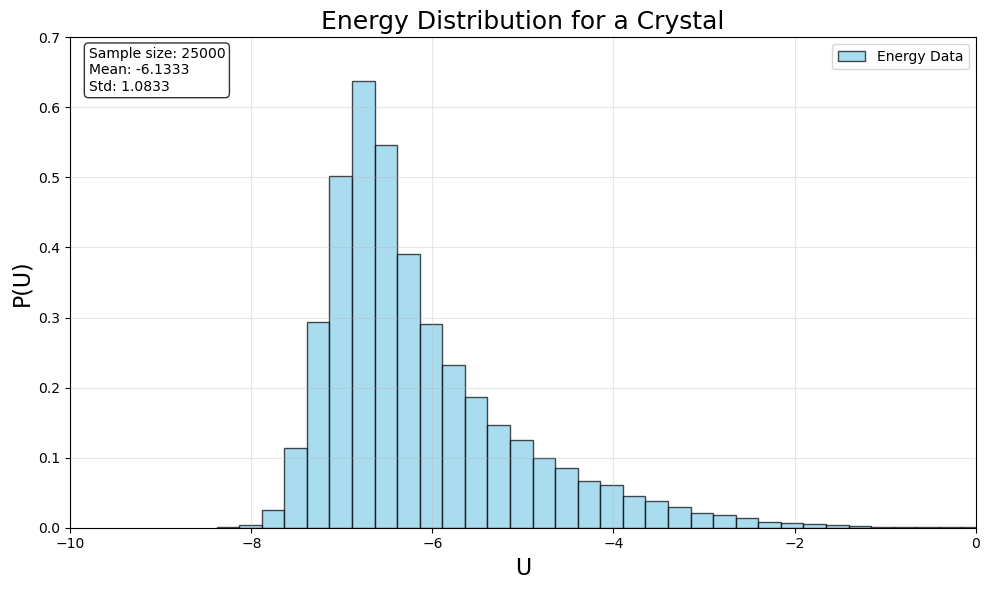}
        \caption{Per-particle $U$ distribution (crystal)}
        \label{fig:hist_lang}
    \end{subfigure}
    \caption[Local energy comparison]{ Single particle potential energy in the supercooled and crystalline states under two thermostatting protocols. (a,b) Spatial distributions of per-particle potential energy for a noisy non-Gaussian (stochastic) thermostat (supercooled) and a Langevin thermostat (crystal). (c,d) Corresponding normalized per-particle potential-energy distributions, highlighting the sharper (narrower) energy distribution in the crystalline state. Notably, at the same nominal temperature the crystalline state exhibits a lower mean potential energy than the supercooled state.}
    \label{fig:potential_4panel}
\end{figure*}

Apart from computing $F(q,t)$, we also examined the  distribution of the local potential energies of the liquid at various degrees of coarse-graining. The local potential energy is defined as follows:
\begin{equation}
U_{j} := \frac{1}{N_{j}} \sum_{i=1}^{N_{j}}\sum_{s=1}^{l} U_{is}( R_{is}).
\end{equation}
Here, $U_{j}$ is the local potential energy of a smaller cell $j$ with $N_j$ being the number of particles in this cell, and $l$ is the number of neighbors of particle $i$. Examples of the resulting single particle potential energy spatial maps and distributions are compiled in Fig.~\ref{fig:potential_4panel}.

The distributions in Fig.~\ref{fig:potential_4panel} show that the
standard deviation of the single particle potential energy is
noticeably larger in the supercooled liquid than in the crystal:
$\sigma_U^{\rm sc}/\sigma_U^{\rm eq} \approx 1.08$ for the
$\overline{A} = 0.09$ system shown.  This $8\%$ excess is far larger
than what the thermal heterogeneity $\overline{A}$ alone can produce,
and its origin lies in the density fluctuations of
$\mathcal{P}_{\rm config}$.  The argument is straightforward.
For a single Cartesian kinetic-energy component
$K_x = \tfrac{1}{2}mv_x^2$, the equilibrium variance is
$\mathrm{Var}(K_x)_{\rm eq} = T^2/2$ (in the Gaussian system,
$\langle v_x^4\rangle = 3\langle v_x^2\rangle^2$).
Applying the law of total variance with $P(T|T')$,
the excess variance in the supercooled state is
$\mathrm{Var}(K_x)_{\rm sc} - \mathrm{Var}(K_x)_{\rm eq}
= \tfrac{3}{4}\overline{A}^2 T^2$, giving a relative excess
standard deviation
$(\sigma_{K_x}^{\rm sc} - \sigma_{K_x}^{\rm eq})/\sigma_{K_x}^{\rm eq}
\approx \tfrac{3}{4}\overline{A}^2 \approx 0.6\%$ for
$\overline{A} = 0.09$.  No assumption about the interaction potential
is needed; the same result follows directly from the measured kurtosis
$\kappa \approx 3\overline{A}^2$ via
$(\sigma_{K_x}^{\rm sc} - \sigma_{K_x}^{\rm eq})/\sigma_{K_x}^{\rm eq}
\approx \kappa/4$.
For the potential energy, the thermal contribution from harmonic
fluctuations about the inherent-structure minima is identical to
the kinetic contribution. Since $U_x = \frac{1}{2}kx^2$ with
$x\sim\mathcal{N}(0,T'/k)$ has exactly the same form as $K_x$
(both are $(T'/2)\cdot Z^2$ with $Z\sim\mathcal{N}(0,1)$),
the law of total variance applied to $P(T|T')$ gives identically:
\begin{eqnarray}
 &&  \frac{\mathrm{Var}(U_x)_{\rm sc} - \mathrm{Var}(U_x)_{\rm eq}}{\mathrm{Var}(U_x)_{\rm eq}}
  = \frac{\mathrm{Var}(K_x)_{\rm sc} - \mathrm{Var}(K_x)_{\rm eq}}{\mathrm{Var}(K_x)_{\rm eq}} \nonumber
\\  && = \tfrac{3}{2}\overline{A}^2 = \frac{\kappa}{2}\approx 1.2\%.
  \label{eq:rel_var_equal}
\end{eqnarray}
This thermal harmonic contribution to the relative variance excess
of $U$ is directly readable from the measured velocity kurtosis
$\kappa$, with no assumption about the interaction potential.
The observed relative variance excess, $(\sigma_U^{\rm sc}/\sigma_U^{\rm eq})^2 - 1 \approx 1.08^2 - 1 \approx 16.6\%$,
is $\sim\!14$ times larger than this thermal prediction of $1.2\%$.
The bulk ($\approx 93\%$) of the single particle potential energy
variance excess therefore arises not from thermal fluctuations about
the inherent-structure minima, but from the \emph{spatial distribution
of the minima themselves} --- i.e.\ from the spread of
$U_{\rm min}$ values across different local regions of the
supercooled liquid encoded in $\mathcal{P}_{\rm config}$.
The steeply density-dependent LJ potential means that regions of
different local density $n'$ (Eq. (\ref{eq:localP})) sit at substantially different
inherent structure energies, contributing a term
$\sim(\partial U_{\rm min}/\partial n')^2\sigma_{n'}^2$
to $\mathrm{Var}(U)_{\rm sc}$ that dominates the thermal term
by an order of magnitude.
The excess $\sigma_U$ is thus a direct signature of the
distribution of particle number densities/inherent structure minima across the spatially
heterogeneous supercooled state.

\section{Conclusion}

The rich spatially nonuniform configuration space landscape of supercooled liquids has been studied for many decades by now. However, the momentum space distributions has been assumed to be canonical. We explored this unquestioned assumption. Our results hint at a possible self-consistent symmetric picture in which both the configurational potential and the kinetic energies differ from those in conventional equilibrium systems. Such deviations cannot be numerically probed with conventional thermostats that, by consruction, are guaranteed to generate Maxwellian velocity distributions. Towards this end, we constructed new thermostats. By systematically comparing continuous equilibrium thermostats with our collision based stochastic schemes, we demonstrated that non-Maxwellian velocity distributions may appear. The excess velocity kurtosis $\kappa$ was introduced as a quantitative dimensionless deviation from Maxwellian statistics. Our studied stochastic collision-based thermostats sustain finite positive $\kappa$, reflecting broadened velocity distributions. These results suggest a consistent link between momentum space broadening and structural evolution. Across extensive ensembles of supercooling runs, the probability of crystallization decreases monotonically with increasing $\kappa$- wider velocity distributions systematically correlate with inhibited nucleation (Section \ref{sec:2.4}).  A non-Gaussian (including heavy tailed) bath generically which yields, as we find, a stationary velocity distribution with finite, nonzero kurtosis, is consistent with a nontrivial fixed-point property of the full coupled dynamics (so long as the supercooled system remains in a steady state prior to crystallization). Our findings of finite kurtosis and nontrivial velocity distributions in supercooled liquids are further consistent with similar trends in other arenas in which stochastic forcing and other effects yield non-Gaussian stationary distributions with persistent higher order moments (including kurtosis), e.g., \cite{Yu2020,Kanazawa2015,Patron2021,Megias2022}. With no drift and leading to stationary averages, the bath may be regarded as part of the larger dynamical system, which self consistently determines the steady state ensemble.
Starting from the single particle velocity distribution as a mixture of Maxwellians, we demonstrated how an effective temperature distribution $P(T|T')$ can be reconstructed from simulation data using numerically stabilized inversion.  
In the supercooled regime, the reconstructed$P(T|T')$ is well described by a Gamma distribution. The reciprocal of the shape parameter ($\overline{A}^2$) quantifies the relative variance of temperature fluctuations and is measurable from the velocity kurtosis via Eq. (\ref{ksim}). The corresponding non-factorizable and hence, as we emphasize and elaborate on, constrained correlated structure of the velocity distribution is discussed in Appendix \ref{app:correlated}.

Most importantly, the high temperature tail area $\mathcal{A}(T'>T_m)$ of the distribution $P$ above the equilibrium melting (or liquidus temperature) tracks, within numerical error, the reciprocal of the shear viscosity. Specifically, as the temperature $T$ drops and the supercooled liquid approaches the glass transition temperature $T_g$, the tail area $\mathcal{A}$ decreases smoothly. Its reciprocal steadily increasingly with the measured viscosity following an apparent VFT like increase. This allows for a direct bridge between microscopic temperature fluctuations and macroscopic transport slowdown without resorting to phenomenological kinetic assumptions.  In strict equilibrium, $\mathcal{A}=0$ and the construction collapses to a single Maxwellian, highlighting that the observed sigmoid crossover between $T_m$ and $T_g$ is intrinsically tied to nonequilibrium ensemble broadening. We find that the simulated systems are consistent with values of $\overline{A}$ with experimental viscosity data of fluids of different types (Fig. \ref{Collapse.}). The simulation viscosities entering this comparison were obtained via the SLLOD algorithm of Appendix \ref{app:sllod}.

Taken together, the inverse Laplace reconstruction of the distribution $P$, its Gamma description, and the associated kurtosis of the studied Lennard-Jones type systems provide a self-consistent interpretation of heterogeneous dynamics in supercooled  liquids. The resulting framework may rationalize behaviors of real (experimental) supercooled fluids. Our approach converts momentum space deviations from Maxwellian statistics into a measurable distribution of effective
temperatures and identifies the width of this distribution as a key control parameter governing relaxation slowdown. The emergence of ever slower sluggish glassy dynamics as the system temperature drops is a consequence of a progressively narrowing support of rare fluctuations that emulate the equilibrium system above its melting temperature $T'=T_{m}$.

On a technical level, the stochastic thermostats introduced in this work, the HTLA and Two-Bath Andersen schemes, provide a principled controllable route to non-Maxwellian velocity statistics in MD simulations. The HTLA thermostat draws post collision relative velocities from a Student-$t$ distribution with tunable tail index. This yields a steady state kurtosis $k = 3(\nu_s-2)/(\nu_s-4)$ for $\nu_s > 4$ degrees of freedom, i.e.,\ an excess kurtosis relative to the Maxwellian (Gaussian) distribution of  $\kappa:= k-3 = 6/(\nu_s-4)$. The Two-Bath thermostat couples particles stochastically to two reservoirs at cold and hot temperatures $T_c$ and $T_h$. For the symmetric case, $p_{\rm hot}=1/2$ with dimensionless contrast $c:=(T_h-T_c)/(T_h+T_c)$, the steady state is a balanced mixture of two Maxwellians whose excess kurtosis is $\kappa(c) = 3c^2$. The agreement with $\kappa\simeq3\overline{A}^{2}$ enables a controlled non-Maxwellian environment. Increasing the contrast parameter broadens the velocity distribution, suppresses crystallization, and increases the structural relaxation time $\langle\tau_\alpha\rangle$. The specific details of our thermostats are given in Appendix~\ref{app:thermostats}. 

A broadened probability distribution $P(T|T')$, and hence a finite initial excess kurtosis, is itself a consequence of supercooling in various situations (see Ref.~\cite{preprint6} focusing on closed systems and Appendix~\ref{rem:drive}).
Once supercooling ceases, a closed system (for which moments of its Hamiltonian trivially remain time translationally invariant) will continue to exhibit a finite standard deviation of its energy density. For general open systems, the reduced probability density relation of Eq. (\ref{eq:localP}) applies. Although a full system equilibrium probability density is trivially time translationally invariant, initial reduced few particle probability densities of an equilibrium form may vary in time, governed by BBGKY type equations. When these reduced equilibrium probability densities $\rho_{{\sf eq.},n}$ are (nearly) stationary, and if the initial excess velocity kurtosis is finite (i.e., if $P$ is not a delta function in the temperature), then the probability density of Eq. (\ref{eq:localP}) will retain this post-supercooling velocity kurtosis for all (respectively, long) times. A plethora of different systems may share identical reduced probability densities, including those of the single particle momentum (or velocity) distribution. There are also non-stationary spatially uniform states with lifetimes much shorter than those of the supercooled state for which initial velocity kurtosis will decay rapidly, e.g., the uniform dilute gas  type system discussed in Appendix \ref{app:boltzmann}. As seen from Liouville's equation, an asymmetry between the configuration and momentum space effective temperature distributions in a liquid will generally give rise to drift of the probability density with the respective contributions being proportional to the mechanical power $\dot K$ and the differences between the kinetic and configuration space inverse temperatures. To close our circle of ideas, to date the focus has been on asymmetric (highly complex and rich) non-uniform spatial configuration and (trivial) uniform momentum space descriptions with standard thermostats not allowing richer kinetic descriptions. We hope that our work will motivate further studies of enabling configuration and momentum space descriptions to be more symmetric as they are at the level of the equations of motion. In particular, simulation protocols that do not force Gaussian velocities, and measurements (e.g., deep inelastic neutron scattering and particle resolved colloidal velocities) that may probe the momentum sector directly. 

\begin{acknowledgments}
This work is dedicated to the memory of our dear colleague, mentor, and friend  Kenneth F. Kelton who is sorely missed.  We are thankful to numerous colleagues with whom we interacted during the course of this work. In particular, we are grateful to K. F. Kelton, Shivaji L. Sondhi, and Gilles Tarjus for many discussions and much encouragement. GT was supported by the McDonnell International Scholar Academy at Washington University in Saint Louis. ZN's work at Oxford and the Laboratoire de Physique Th\'eorique de la Mati\`ere Condens\'ee (LPTMC) were, respectively, supported by a Leverhulme Trust International Professorship (S. L. Sondhi) grant (number LIP-2020-014) and by the CNRS.
\end{acknowledgments}

\appendix

\section{Reduced probability densities of metastable states: review}
\label{app:theorem}

\noindent In order to make notions that motivated the current work more self-contained, we review some of these elements.

\subsection{Stationary reduced probability densities}

The central result of Ref.~\cite{Nussinov2024}, complementing the framework~\cite{grant16,grant15,preprint2,preprint6,Planck22,book_TOP}, that led to the prediction of non-Maxwellian velocity distributions links the reduced few body
($n = \mathcal{O}(1)$) probability densities of macroscopic ($N \gg 1$)
metastable systems to their thermal equilibrium counterparts.  Under
ensemble equivalence (which holds for conventional disorder-free
Hamiltonians governing supercooled liquids and their equilibrium
counterparts), the reduced $n$-particle probability density of a
stationary state at inverse temperature $\beta$, chemical potential
$\mu, \ldots$ is
\begin{eqnarray}
\label{central-theorem}
&& \rho_{{\sf stationary},n}(\beta, \mu, \ldots) = \nonumber \\
&& \int d\beta' d \mu' \ldots \Big( P(\beta,  \mu, \ldots| \beta', \mu', \ldots )
\nonumber \\
&& \times \rho_{{\sf eq.},n}(\beta',  \mu', \ldots) \Big) \nonumber
\\   && +
\int_{\cal{CO}} d\epsilon' d {\tilde{{\sf n}}}'  \cdots
\Big(P(\epsilon, {\tilde{{\sf n}}},  \ldots| \epsilon', {\tilde{{\sf n}}}'
,\ldots) \nonumber \\
&& \times \rho_{{\sf eq.},n} (\epsilon', {\tilde{{\sf n}}}',  \ldots) \Big).
\end{eqnarray}
It should be emphasized that this is not if and only if condition. That is, not all reduced probabilities of the weighted form of Eq. (\ref{central-theorem}) need to be stationary. Rather, what Ref. ~\cite{Nussinov2024} established is that all stationary reduced probability densities must be of the above form. Eq. (\ref{central-theorem}) is an extended and slightly more precise form of Eq. (\ref{eq:localP}) when there is a region of equilibrium phase coexistence ${\cal{CO}}$. Here, $\rho_n := {\sf Tr}_{n+1,\ldots,N}(\rho_\Lambda)$ is the partial
trace of the full many body probability density. A simplified form of
this equation, omitting the coexistence term, is stated in the main
text (Section~II); the present Eq.~(\ref{central-theorem}) is more
complete in that it explicitly retains the contribution from the
phase-coexistence region.

Stationarity of the long time average
$\rho^* := \lim_{\tau\to\infty}\frac{1}{\tau}
\int_t^{t+\tau}\rho_\Lambda(t')\,dt'$
implies $[\rho^*,H]=0$ ( in the quantum arena) or $\{\rho^*,H\}_{P.B.}=0$ (classical).
The most general solution to these commutator or Poisson bracket conditions 
is a weighted sum of projection operators onto states of fixed energy $E'$ and conserved quantities
$\{W'_\alpha\}$:
\begin{equation}
\label{PEW}
{\mathbb{P}}_{E',\{W'_{\alpha}\}} = |E', \{W'_{\alpha}\}\rangle\langle E',\{W'_{\alpha}\}|.
\end{equation}
Whenever ensemble equivalence holds and $\mathcal{P}$ is smooth, this
yields Eq.~\eqref{central-theorem} with the weight
$\mathcal{P}(E',\{W'_\alpha\})$ identified with
$P(\beta,\ldots|\beta',\ldots)$.

Several remarks are in order:
\begin{enumerate}\setlength\itemsep{2pt}
\item The linear map of Eq.~\eqref{central-theorem} applies
  \emph{only} to stationary $\rho_{{\sf stationary},n}$. Strictly speaking, it cannot
  link equilibrium densities to arbitrary time evolving probability
  densities.
\item For a bona fide equilibrium system,
  $P(\beta,\mu,\ldots|\beta',\mu',\ldots) =
  \delta(\beta'-\beta)\delta(\mu'-\mu)\cdots$, recovering the standard
  result.
\item Eq.~\eqref{central-theorem} should also apply to the putative ideal glass
  (conjectured to be permanently quiescent  ~\cite{Cavagna,BB})
  whenever ensemble equivalence holds.
\end{enumerate}

In general, although the system may have a fixed particle number or energy,
whenever the reduced few body probability density is stationary the system
effectively lies in the convex hull of equilibrium states~\cite{Nussinov2024}
and samples equilibrium states of different energy densities, particle
densities, and any other conserved quantities, as in
Eq.~\eqref{central-theorem} of the macroscopic system.
Conceptually, Eq.~\eqref{central-theorem} bears some similarity to, and
may be further refined by, the ergodic decomposition theorem whenever the
system has disjoint ergodic sectors. Unlike the usual textbook realizations
of the ergodic theorem, the ergodic sectors in Eq.~\eqref{central-theorem}
are effectively labeled by different intensive state variables and other
constants of motion densities. Ref.~\cite{preprint2} employed the above
notion, taking the equilibrium particle density as a state variable, to
examine the jamming transition. Since the particle density of the crystal
and that of the supercooled liquid differ, contributions from the
coexistence region (the $\int_{\cal{CO}}$ term in
Eq.~\eqref{central-theorem}) are in general non negligible and must be
retained in a complete treatment.

\subsection{Near stationarity of the supercooled liquid}

In the metastable supercooled liquid at $T > T_g$, local few body
observables are nearly (but not perfectly) stationary, since
crystallization occurs after a finite time $\tau_{\sf xtal}$.  The
ratio of microscopic to crystallization timescales,
\begin{eqnarray}
\label{scales-eq}
    10^{-10} \gtrsim \tau_{\sf micro}/\tau_{\sf xtal} \gtrsim 10^{-17},
\end{eqnarray}
ensures that corrections to Eq.~\eqref{central-theorem} are
exceptionally small~\cite{Nussinov2024}.  These bounds include metallic glass formers (where the crystallization times $\tau_{\sf xtal}\gtrsim 10^{-3}$~s) to silicate and
organic systems ($\tau_{\sf xtal}\gtrsim 10^4$~s just above
$T_g$)~\cite{Kelton_review}.

\begin{tcolorbox}[enhanced, colback=yellow!10!white, colframe=black,
  boxrule=0.7pt, arc=2pt, left=8pt, right=8pt, top=6pt, bottom=6pt]
Since local few body expectation values in the metastable supercooled
fluid differ from those in the equilibrium liquid or crystal, the
distribution $P$ in Eq.~\eqref{central-theorem} must differ from the
equilibrium delta-function form.  $P$ must therefore have a
\textbf{finite standard deviation} in at least one state variable.
\end{tcolorbox}

\begin{remark}[time averaged versus instantaneous correlators]
When the probability densities $\rho_\Lambda(t')$ are not completely
stationary, connected correlation functions computed with their
time average can be spatially longer ranged than their instantaneous
counterparts.  To see this precisely, let $A$ and $B$ be local
observables at spatially separated points, and let
$\overline{X(t')} = \frac{1}{\tau}\int_t^{t+\tau} X(t')\,dt'$
denote the time average computed with the uniform distribution
$P_\tau(t') = \frac{1}{\tau}\Theta(t+\tau-t')\Theta(t'-t)$.
Suppose that at every instant $t'$ the equilibrium probability density
$\rho_\Lambda(t')$ has only local spatial correlations, so that
$\langle A(t')B(t')\rangle - \langle A(t')\rangle\langle B(t')\rangle
\to 0$ for $A$ and $B$ far apart.
Nevertheless, the time averaged covariance
\begin{equation}
  \overline{\langle A(t')B(t')\rangle}
  - \overline{\langle A(t')\rangle}\;\overline{\langle B(t')\rangle}
  \label{eq:time_avg_covar}
\end{equation}
need not vanish, because the time average mixes instantaneous states
at different $t'$, and cross-correlations between $\langle A(t')\rangle$
evaluated at one time and $\langle B(t'')\rangle$ at another can
survive the averaging even when the instantaneous connected correlator
is short-ranged.  This is precisely why \emph{only empirically measured
local} finite-particle expectation values need be nearly static in the
supercooled state: global stationarity of $\rho_\Lambda$ is not
required, and long range correlations in time averaged quantities do
not contradict the short-range nature of the instantaneous equilibrium
densities.

Finally, we note that the reduced $n$-particle probability density of
Eq.~\eqref{central-theorem} can be amended by an additive contribution
$\mathrm{Tr}_{n+1,\ldots,N}(f_\Lambda)$, where $f_\Lambda$ is a
correction to the full many body density that keeps the reduced
probability density positive semidefinite while leaving the
near stationarity condition intact.
\end{remark}

\subsection{An origin of finite width of $P(T|T')$}
\label{rem:drive}
We now outline a mechanism~\cite{preprint6} by which the distribution $P(T|T')$ acquires a finite width.

Let $\{\phi_i\}_{i=1}^N$ denote compact local fields (e.g.\ local energy or order parameter densities) at $N$ spatial sites, governed by a local Hamiltonian
$H_0[\{\phi_i\}] = \sum_i\bigl[\tfrac{r}{2}\phi_i^2 + \tfrac{u}{4}\phi_i^4 + \cdots\bigr]$.
During rapid cooling, the external bath couples linearly to all sites
through a common fluctuating field $h(t)$. This is modeled by the term
\begin{eqnarray}
  H_{\rm drive} = -h(t)\sum_i \phi_i,
  \qquad \nonumber
  \\ 
  \langle h(t)\rangle = 0,\quad  \langle h(t)h(t')\rangle = D\,\delta(t-t').
  \label{eq:Hdrive}
\end{eqnarray}
The coupling must be that to an {\it extensive} fraction of sites in order to change the global energy density $\bar{E}=(1/N)\sum_i\epsilon_i$ at a finite rate.

We next integrate out the fluctuating field $h(t)$. 
Rather explicitly, the path integral over $\{\phi_i\}$ and $h$ is
$Z = \int \mathcal{D}[\{\phi_i\}]\,\mathcal{D}[h]\,e^{-S_0-S_{\rm drive}}$,
with $S_{\rm drive} = \int dt\bigl[-h\sum_i\phi_i + h^2/(2D)\bigr]$.
Trivially completing the square, 
\begin{equation}
  -h\sum_i\phi_i + \frac{h^2}{2D}
  = \frac{1}{2D}\Bigl(h - D\sum_i\phi_i\Bigr)^2
  - \frac{D}{2}\Bigl(\sum_i\phi_i\Bigr)^2,
\end{equation}
and integrating out $h$ leads to the effective action
\begin{equation}
  S_{\rm eff} = S_0 - \frac{D}{2}\int dt\,\sum_{i,j}\phi_i(t)\phi_j(t).
  \label{eq:Seff}
\end{equation}
The integrated out drive $h$ generates an \emph{all-to-all coupling} $-\tfrac{D}{2}\phi_i\phi_j$ between every pair of sites, including $i\neq j$. The all-to-all coupling in Eq. (\ref{eq:Seff}) yields a positive covariance between any pair of sites, 
\begin{equation}
  \mathrm{Cov}(\phi_i,\phi_j) = D\chi^2\tau_{\rm drive} = \mathcal{O}(1).
  \label{eq:cov_drive}
\end{equation}
Here, $\chi$ is the local susceptibility and $\tau_{\rm drive}$ is the duration of the cooling. The variance of the global intensive average $\bar\Phi = N^{-1}\sum_i\phi_i$ is then system size independent,
\begin{eqnarray}
  \mathrm{Var}(\bar\Phi)
  = \frac{1}{N^2}\sum_{i,j}\mathrm{Cov}(\phi_i,\phi_j) \nonumber
\\   = \frac{N^2}{N^2}\,D\chi^2\tau_{\rm drive}
  = \mathcal{O}(1).
  \label{eq:var_global}
\end{eqnarray}
The above result is diametrically opposite to the case of independent fluctuations. For the latter case, in the thermodynamic limit, $\mathrm{Var}(\bar\Phi)_{\rm indep} = \sigma_{\rm local}^2/N\to 0$. The compactness of the fields (the $u\phi^4$ term) ensures $\sigma_\epsilon = \mathcal{O}(1)$ remains finite rather than diverging with $\tau_{\rm drive}$.

Once the common drive ceases to act, each site couples to its own independent local noise $\xi_i(t)$ with $\langle\xi_i(t)\xi_j(t')\rangle = 2T\delta_{ij}\delta(t-t')$. The intersite covariance then decays,
\begin{equation}
  \frac{d}{dt}\langle\phi_i\phi_j\rangle
  \approx -(\gamma_i+\gamma_j)\langle\phi_i\phi_j\rangle,
  \qquad i\neq j,
\end{equation}
exponentially on local relaxation timescale. The all-to-all coupling in Eq.~\eqref{eq:Seff} vanishes and $\mathrm{Cov}(\phi_i,\phi_j)\to 0$ for $i\neq j$. Consequently,
\begin{equation}
  \mathrm{Var}(\bar\Phi)_{\rm after} = \frac{1}{N}\mathrm{Var}(\phi_i) = \mathcal{O}(1/N) \to 0.
\end{equation}
A finite $\mathcal{O}(1)$ standard deviation of the global average in the stationary state will require long range correlations that are no longer present \cite{preprint6}.

In the content of rapid cooling, each local field $\phi_i$ acted on by the drive was displaced to a value encoding a local effective temperature $T'_i$. Once the drive (cooling) ceases, the system cannot veer towards its true equilibrium crystalline state on short times.  Just below the melting temperature, critical nuclei are large and rare, yielding low nucleation rate. At the other (low temperature) limit of the supercooled system, the viscosity is enormous and the dynamics are arrested, so even though crystalline nuclei may form they cannot grow in size.  Held between these two pincers (few nuclei near melting and frozen dynamics at low temperature), the distribution of the fields $\{\phi_i\}$ may become arrested on exceptionally long timescales. The distribution of $\{T'_i\}$ across different regions of the system is of width
\begin{equation}
  \sigma(T') 
  = \sqrt{\mathrm{Var}(\{T'_i\})}
  = \mathcal{O}(\Abar\, T).
  \label{eq:sigma_spatial}
\end{equation}
That is, the standard deviation of the fields is system size independent, $\mathcal{O}(1)$. It is precisely this system size independent width of $T'$ in the \emph{reduced few body local probability density} (borne by a spatial variation of effective local temperatures across the system), that constitutes the finite width $P(T|T')$ of Eq.~\eqref{central-theorem}. This wide (i.e., non delta function) distribution $P$ gives rise to the non-Maxwellian velocity statistics observed in the present work. 

Inequalities for maintaining equilibrium systems can be arrived by applying this logic in reverse. That is, demanding that $P$ have a vanishing width so that the system is in equilibrium leads to bounds on the maximal rates of change of observables \cite{preprint6,Planck22}. 

\subsection{Viscosity from local observables}
\label{app:viscosity_local}

Setting the observable $Q$ in Eq.~\eqref{central-theorem} to the
terminal velocity of a sphere falling through the liquid (Stokes
equation), and noting that equilibrium flow occurs only above the
liquidus temperature $T_l$, gives~\cite{book_TOP,grant16,grant15,preprint2,Nussinov2024}:
\begin{eqnarray}
\label{viscosityfromv}
\eta(T) \simeq \frac{\eta_{\sf eq.}(T^+_{l})}{\int_{T_{l}}^{\infty} dT' P(T|T')},
\end{eqnarray}
which with a Gaussian $P(T|T')$ of width $\overline{A}T$ yields the erfc collapse
of Eq.~\eqref{eq:visc1}.

\begin{tcolorbox}[enhanced, colback=yellow!10!white, colframe=black,
  boxrule=0.7pt, arc=2pt, left=8pt, right=8pt, top=6pt, bottom=6pt]
A \textbf{unique} function $P$ should consistently provide predictions
for \textit{all} local few body observables of a given fluid.
Known singularities of equilibrium phase transitions will be
``smeared out'' by a $P$ that is not a delta function, yet the
equilibrium melting temperature still plays a prominent role ---
underlying the appearance of $T_l$ in the viscosity collapse.
\end{tcolorbox}

\section{Kurtosis decay in a dilute gas with elastic binary  collisions: Boltzmann equation analysis}
\label{app:boltzmann}

In the Appendix, we study the decay of an initial velocity kurtosis for a spatially homogeneous (dilute gas) system within the framework of the Boltzmann equation,
\begin{equation}
  \partial_t f(\vec{v},t) = C[f,f].
\end{equation}
Here, $C[f,f]$ is the elastic binary collision operator conserving
particle number, momentum, and kinetic energy. In what follows, we introduce shorthands for the velocity moments
$M_n = \int d\vec{v}\, |\vec{v}|^n f(\vec{v},t)$.  Particle number, momentum, and energy conservation imply that  $\dot{M}_0 = \dot{M}_2 = 0$.  The per component excess kurtosis $\kappa=\langle v_x^4\rangle/\langle v_x^2\rangle^2-3$ that we used throughout the current work is, for an isotropic $f$,
\begin{equation}
  \kt = \frac{9}{5}\,\frac{M_4}{M_2^2} - 3.
\end{equation}
As it must, $\kt$ vanishes for the Maxwell--Boltzmann distribution (where $M_4/M_2^2=5/3$). Since $M_2$ is conserved, the evolution of the excess kurtosis is determined entirely by $\dot{M}_4$.

Using the standard symmetrized form of $C[f,f]$ and introducing
center of mass and relative velocities, $\vec{V} := (\vec{v}_1
+ \vec{v}_2)/2$, $\vec{g} := \vec{v}_1 - \vec{v}_2$, with
$|\vec{g}'|=|\vec{g}|$ following the collision, 
with $\theta$ the angle between $\vec{g}'$ and $\vec{g}$, and a solid angle $d\Omega$ integral over the direction of $\vec{g}'$,
we have that
\begin{equation}
  \frac{dM_4}{dt}
  = \frac{1}{4}\int d\vec{v}_1\,d\vec{v}_2\,d\Omega\;
    B(g,\theta)\,f_1 f_2
    \bigl(v_1'^{\,4}+v_2'^{\,4}-v_1^4-v_2^4\bigr).
    \label{eq:M4dot_full}
\end{equation}
A direct calculation yields 
\begin{equation}
  v_1'^{\,4}+v_2'^{\,4}-v_1^4-v_2^4
  \;=\;
  2\bigl[(\vec{V}\cdot\vec{g}')^2 - (\vec{V}\cdot\vec{g})^2\bigr].
  \label{eq:diff_exact}
\end{equation}
For Maxwellian particles, $B(g,\theta)=B(\theta)$ is independent of
$g=|\vec{g}|$. Expressing the deviation of the single particle velocity distribution from Maxwellian (M) equilibrium as
$f(\vec{v},t)=f_{\mathrm{M}}(\vec{v})\,[1+h(\vec{v},t)]$ and
linearizing $C[f,f]$ about $f_{\mathrm{M}}$, the perturbation evolves as
$\partial_t h=\mathcal{L}h$. The linearized collision operator $\mathcal{L}h$ is
defined exactly by
\begin{eqnarray}
  (\mathcal{L}h)(\vec{v}_1)
  &:=&\int d\vec{v}_2\int d\Omega\;B(\theta) f_{\mathrm{M}}(\vec{v}_2)\,  
  \\ 
&&  \times  \bigl[\,h(\vec{v}_1')+h(\vec{v}_2')-h(\vec{v}_1)-h(\vec{v}_2)\,\bigr]. \nonumber
  \label{eq:Llin}
\end{eqnarray}
The two gain terms
$h(\vec{v}_{1,2}')$ and two loss terms $h(\vec{v}_{1,2})$ arise
because $f_{\mathrm{M}}(\vec{v}_1')f_{\mathrm{M}}(\vec{v}_2')
=f_{\mathrm{M}}(\vec{v}_1)f_{\mathrm{M}}(\vec{v}_2)$ by energy
conservation. The operator $\mathcal{L}$ is self adjoint and negative
semidefinite with respect to the weighted inner product $\langle
g,h\rangle=\int d\vec{v}\,f_{\mathrm{M}}(\vec{v})\,g(\vec{v})
h(\vec{v})$, with kernel spanned by the collision invariants
$\{1,\vec{v},\vec{v}^2\}$.  For Maxwellian particles, it enjoys an eigenfunction expansion in Sonine (associated Laguerre) polynomials \cite{Bobylev1976,Cercignani1988}.  The fourth order Sonine polynomial,
\begin{equation}
  S_2^{(1/2)}(c^2) \;\propto\; c^4 - 5c^2 + \tfrac{15}{4},
  \quad c^2 = |\vec{v}|^2/(2T),
\end{equation}
is an eigenfunction of $\mathcal{L}$ of eigenvalue $-\nu_4 < 0$.
Here, 
\begin{equation}
  \nu_4 = n \int_0^\pi B(\theta)(1-\cos^4\theta)\sin\theta\,d\theta
  \;\sim\; \frac{1}{\tcoll},
  \label{eq:nu4}
\end{equation}
with $\tcoll$ denoting the microsopic collision time scale. Since the excess kurtosis is proportional to the coefficient of this
eigenfunction in the expansion of $f - f_{\mathrm{M}}$, it satisfies
\begin{equation}
    \frac{d(\kt)}{dt} = -\nu_4\,\kt,
    \qquad
    \kt = \kinit\,e^{-\nu_4 t}.
  \label{eq:kappa_normal}
\end{equation}
The excess kurtosis $\kappa$ is thus not a hydrodynamic mode. Rather, in this system, $\kappa$ decays on the said collision timescale $\tcoll \sim \nu_4^{-1}$, faster than any
hydrodynamic (sound/diffusion) mode, and independently of the
hydrodynamic fields.

The above simple calculation was for an idealized homogeneous system with elastic binary collisions. Supercooled liquids are far more complex (and nonuniform) than such systems. For the supercooled state, the theorem of Ref.~\cite{Nussinov2024}
establishes that the single particle velocity distribution is a weighted sum of Maxwellian distributions at different effective
temperatures $T'$, drawn from a distribution $P(T|T')$ of nonzero
width. For the Gaussian ansatz with dimensionless width $\Abar$,
the steady state excess kurtosis follows from the moment identity of Eq.~(\ref{eq:kappa_Abar}),
\begin{equation}
  \kss = 3\Abar^2,
  \label{eq:kss_A}
\end{equation}
where $\Abar$ also governs the erfc viscosity
formula of Eq. (\ref{eq:visc1})~\cite{grant15,Nussinov2024},
which fits the viscosity of 45 supercooled liquids over 16 decades
with $\Abar \simeq 0.08 \pm 0.03$~\cite{grant15}.

For more realistic interactions (e.g.\ hard spheres), the eigenvalue
$\nu_4$ acquires velocity dependence and nonlinear corrections appear,
\begin{equation}
  \frac{d(\kt)}{dt} = -\nu_4\,\kt + \nu_2\,\kappa^2(t) + \cdots,
\end{equation}
but the dominant term remains $-\nu_4\,\kt$ for small $\kt$.



\section{Thermostat Descriptions}
\label{app:thermostats}

\subsection{Conventional Thermostats}

\subsubsection{Nos\'{e}--Hoover}
    The Nos\'e--Hoover thermostat~\cite{Hoover1985} introduces a deterministic thermal reservoir through a
    dynamically evolving friction coefficient that regulates the kinetic temperature.
    In discrete time, the particle velocities are updated according to
    \begin{equation}
        \vec{v}_{i,n+1} =
        \vec{v}_{i,n}
        + \Delta t \left(
        \frac{\vec{F}_{i,n}}{m_i}
        - \zeta_n\,\vec{v}_{i,n}
        \right),
    \end{equation}
    where the thermostat variable evolves as
    \begin{equation}
        \zeta_{n+1} =
        \zeta_n
        + \frac{\Delta t}{Q}
        \left(
        \frac{T_n}{T_{\text{bath}}} - 1
        \right),
    \end{equation}
    with $Q$ the effective thermostat mass. In the long time limit, this scheme
    samples the canonical ensemble with a Maxwellian velocity distribution.

\subsubsection{Berendsen}
    The Berendsen thermostat~\cite{berendsen1984molecular} regulates the system temperature through global velocity
    rescaling:
    \begin{equation}
        \vec{v}_{i,n+1} =
        \vec{v}_{i,n}
        \left(
        1 + \frac{\Delta t}{\tau}
        \left(
        \frac{T_{\text{bath}}}{T_n} - 1
        \right)
        \right)^{1/2}.
    \end{equation}
    This thermostat is computationally efficient but does not generate a true canonical ensemble, and frequently samples velocity distributions with negative $\kappa$, narrower than Maxwell-Boltzmann.

\subsubsection{Langevin}
The Langevin thermostat~\cite{AllenTildesley2017} models the thermal bath through a combination of viscous
damping and stochastic forcing:
\begin{equation}
\begin{split}
\vec{v}_{i,n+1}
&= \vec{v}_{i,n}
+ \Delta t
\left(
\frac{\vec{F}_{i,n}}{m_i}
- \gamma\,\vec{v}_{i,n}
\right)
+ \sqrt{\frac{2\gamma T_{\text{bath}}\Delta t}{m_i}}\,
\vec{\eta}_{i,n},
\end{split}
\end{equation}
where $\gamma$ is the friction coefficient and $\vec{\eta}_{i,n}$ is a vector of
independent Gaussian random variables with zero mean and unit variance. Langevin dynamics is common practice to achieve the canonical Maxwell-Boltzmann distribution.

\subsection{Non-Equilibrium Thermostats}

\subsubsection{Heavy-Tailed Lowe--Andersen (HTLA) Thermostat}

The HTLA thermostat generalizes the conventional
stochastic Andersen/Lowe~\cite{Andersen_thermostat} collision scheme by replacing the Maxwell--Boltzmann
post collision velocity sampling with a heavy-tailed kernel.

Collision pairs $(i,j)$ are selected
with probability $P_{\mathrm{coll}} = 1 - e^{-\nu \Delta t}$.
For each selected pair, only the relative velocity projected along the
interparticle axis direction,
$g_{ij}^{\parallel} = (\vec{v}_i - \vec{v}_j)\cdot \hat{{r}}_{ij}$,
is resampled. In HTLA, the Gaussian kernel is replaced by a Student-$t$ distribution:
\begin{equation}
P_{\mathrm{HTLA}}(g^{\parallel}) =
\frac{\Gamma\!\left(\frac{\nu_s+1}{2}\right)}
{\sqrt{\nu_s \pi}\,\Gamma\!\left(\frac{\nu_s}{2}\right)\sigma}
\left[
1 + \frac{(g^{\parallel})^2}{\nu_s \sigma^2}
\right]^{-\frac{\nu_s+1}{2}},
\end{equation}
where $\nu_s$ denotes the number of degrees of freedom (the tail index) and $\sigma^2 = T / \mu$ sets the thermal scale. The tail index maps to a control parameter $A$ via $\nu_s = 2/A^2$, so that the kurtosis of the HTLA distribution is
\begin{equation}
\kappa =
\begin{cases}
3(\nu_s - 2)/(\nu_s - 4), & \nu_s > 4, \\
\infty, & \nu_s \le 4.
\end{cases}
\end{equation}
Since only relative velocities are redrawn, linear momentum is conserved pairwise.

\subsubsection{Two-Bath Andersen Thermostat}

The Two-Bath Andersen thermostat extends the conventional Andersen scheme by
coupling particles to two distinct thermal reservoirs at temperatures $T_c$ and $T_h$.
When a collision event occurs for particle $i$ with probability $P_{\mathrm{coll}} = 1 - e^{-\nu \Delta t}$, the hot bath is selected with probability $p_{\mathrm{hot}}$. The post collision velocity is reassigned from the Maxwell--Boltzmann
distribution corresponding to the chosen bath. The unconditional distribution at collision events is therefore a mixture:
\begin{equation}
P(\vec{v})
=
(1 - p_{\mathrm{hot}}) P(\vec{v} \mid T_c)
+
p_{\mathrm{hot}} P(\vec{v} \mid T_h).
\end{equation}
The kurtosis of a single velocity component is
\begin{equation}
\kappa
=
3\,
\frac{
(1 - p_{\mathrm{hot}}) T_c^2
+
p_{\mathrm{hot}} T_h^2
}{
\left[
(1 - p_{\mathrm{hot}}) T_c
+
p_{\mathrm{hot}} T_h
\right]^2
} - 3.
\end{equation}
For the symmetric case $p_{\mathrm{hot}} = \tfrac12$ with parameterization
$T_h = T_0(1+c)$, $T_c = T_0(1-c)$:
\begin{equation}
\kappa(c)
=
3(1 + c^2) - 3 = 3c^2.
\end{equation}
Thus increasing temperature separation produces systematic
heavy-tailed broadening relative to Maxwellian statistics.

\subsubsection{Stochastic Elastic-Collision Thermostat}

We also employ a custom-designed stochastic thermostat that models central,
three-dimensional elastic collisions between MD
particles of mass $M$ and background gas particles of mass $m$.
The post collision velocity of the MD particle is
\begin{equation}
\vec{v}_{n+1}
=
\frac{M - m}{M + m} \vec{v}_n
+
\frac{2m}{M + m} \vec{v}_s,
\label{eq:elastic_update}
\end{equation}
where $\vec{v}_s$ is sampled from a Maxwell--Boltzmann distribution at bath temperature $T_b$. This thermostat provides a physically
transparent mechanism for modeling thermal exchange while preserving microscopic conservation laws at each collision event.

\subsubsection{Other Thermostats}

Additional variants explored include MD-simulated gas-liquid collisions (where an ideal gas is spawned alongside the MD liquid and elastic collisions occur at designated contact distances) and a Randomized Berendsen thermostat in which the coupling parameter is drawn randomly:
\begin{equation}
 \vec{v}_{n+1}=\frac{\vec{v}_n}{|\vec{v}_n|} \left(|\vec{v}_n|+R \left(\sqrt{2T/3m}-|\vec{v}_n|\right)\right),
\end{equation}
where $R\ll 1$ is a randomly chosen small coupling parameter. These variants were found to be less numerically tractable on feasible MD timescales.

\subsection{Drawbacks of Conventional Thermostats}

The Nos\'e--Hoover thermostat imposes a Maxwell--Boltzmann velocity distribution by design. By enforcing this distribution deterministically, it can distort the natural evolution of dynamical observables in non-ergodic or highly heterogeneous states.

The Berendsen thermostat does not generate a true canonical ensemble. Instead of sampling from the correct Maxwell--Boltzmann distribution, it suppresses the natural fluctuations in kinetic energy, artificially narrowing the velocity distribution and distorting fluctuation-derived properties.

The Langevin thermostat enforces canonical sampling by augmenting the equations of motion with viscous damping and stochastic noise satisfying the fluctuation--dissipation relation. While this guarantees convergence to Maxwell--Boltzmann in equilibrium, the explicit stochastic forcing can significantly perturb underlying microscopic dynamics, particularly for slow collective motion and long lived dynamical correlations characteristic of glass-forming systems.

Regardless of these factors, the use of a thermostat remains a necessity for sizable MD systems, as numerical integration errors lead to a gradual increase in the system's total energy manifesting as unphysical heating over time~\cite{FrenkelSmit2002}.

\section{Non-Gaussian Dynamics, Dynamical Heterogeneity, and Crystallization}
\label{app:nongaussian}

To better place our results in context, in this appendix, we all too briefly review well known  elements, e.g.,  \cite{DH1,DH2,DH3,DH4,DH5,dh10,volynes,dh11,dh12,PhysRevLett.132.258201,doi:10.1021/acs.jpcb.5b08912,Kelton_review,BB,Chauduri,Bhowmik,Rahman1964,Donati1999,Fredrickson1984,JackleEisinger1991,Chandler2010,Mirigian2013unified,Palmer1984,Berthier2021,keys}. 

\subsection{Non-Gaussian observables}
The excess (single Cartesian component or other generalized coordinate) velocity kurtosis,
is the central quantity analyzed in the current work. 
Other metrics of non-Gaussian statistics have been heavily used in earlier
studies of supercooled liquids. These include, in particular, the
\emph{displacement} non-Gaussian parameter, introduced by
Rahman~\cite{Rahman1964},
\begin{equation}
  \alpha_2(t) = \frac{3\langle (\vec{r}^2(t))^2\rangle}{5\langle \vec{r}^2(t)\rangle^2} - 1.
  \label{eq:alpha2_def}
\end{equation}
The prefactor $3/5$ in $\alpha_2$ is, in general spatial dimension $d \neq 3$, replaced by $d/(d+2)$. 

\subsection{Dynamical heterogeneity}

As noted in the main text, in supercooled liquids, dynamics may become spatially heterogeneous \cite{DH1,DH2,DH3,DH4,DH5,dh10,volynes,dh11,dh12}. Herein, fast and slow regions coexist and relaxation occurs via cooperative rearrangements. Notably, displacement distributions develop non-Gaussian
tails~\cite{Donati1999,BB}. The focus of all earlier studies has not been on velocities (which until the current work have been invariably assumed to be Maxwellian) but rather on relaxation rates. 

A key empirical fact is that $\alpha_2(t)$ peaks at the structural
($\alpha$-)relaxation time $\tau_\alpha$~\cite{Kob-Andersen,Donati1999}. The peak of $\alpha_2(t)$
becomes larger and monotonically shifts to longer times as the temperature drops. This tend tracks the
growth of cooperative rearranging regions.

\subsection{Kinetically constrained models}

Kinetically constrained models (KCMs) \cite{Mirigian2013unified, Palmer1984,Berthier2021,keys}, including the
Fredrickson-Andersen (FA) and East models~\cite{Fredrickson1984,JackleEisinger1991,Chandler2010}, capture these features through local
mobility constraints. Herein, dynamics are facilitated by neighboring excitations. The resulting relaxation is hierarchical and spatially heterogeneous. The structural relaxation time grows in a super-Arrhenius fashion. Within the East model, $\tau_\alpha \sim \exp(C/T^2)$  with $C$ being a constant \cite{keys}.

Far more general than the domain of kinetically constrained models alone, the distribution of local relaxation times may be broad. This may lead to stretched exponential relaxation~\cite{Palmer1984} and consequently yield large values of $\alpha_2(t)$.

\subsection{Relation between $\alpha_2(t)$ and relaxation}

The non-Gaussian parameter is related to the variance of the local
diffusivity $D$~\cite{Berthier2004},
\begin{equation}
  \alpha_2(t) \;\sim\; \frac{\mathrm{Var}(D)}{\langle D\rangle^2}.
\end{equation}
As a general rule of thumb trend, increasing non-Gaussianity $\alpha_2$ appears in tandem with stronger dynamical heterogeneity.

\subsection{Connection to the crystallization time $\tau_{\mathrm{xtal}}$}

The classical nucleation rate is governed by~\cite{TurnbullFisher1949} $
  \tau_{\mathrm{xtal}} \;\sim\; \tau_0\,e^{\Delta F^*/T}$, 
where $\Delta F^*$ is the nucleation barrier and $\tau_0$ is an 
attempt timescale. In a supercooled liquid, nucleation requires
collective rearrangements. In other words, $\tau_0 \sim \tau_\alpha$~\cite{Wedekind2008} and 
\begin{equation}
  \tau_{\mathrm{xtal}} \;\sim\; \tau_\alpha\,e^{\Delta F^*/T}.
  \label{eq:txtal_talpha}
\end{equation}
Since $\alpha_2(t)$ peaks at times $t \sim \tau_\alpha$, its lifetime and
magnitude are indirect proxies for $\tau_\alpha$ and hence for
$\tau_{\mathrm{xtal}}$.

\subsection{Competing effects of non-Gaussianity}

Non-Gaussianity has two opposing consequences for crystallization.
\begin{enumerate}\setlength\itemsep{2pt}
  \item Large $\alpha_2$ signals slow, heterogeneous dynamics
    $\Rightarrow \tau_\alpha\uparrow \Rightarrow \tau_{\mathrm{xtal}}\uparrow$.
  \item Concomitantly, however, similar to the discussion in the main paper (Figs. \ref{fig:InvTailArea_VFTfit} and \ref{fig:PT_and_tail_area}), fat tails may imply an enhanced tail of fast particles
    $\Rightarrow$ locally enhanced barrier crossing
    $\Rightarrow \tau_{\mathrm{xtal}}\downarrow$.
\end{enumerate}
Thus $\alpha_2$ correlates with $\tau_{\mathrm{xtal}}$ but does not
uniquely determine it.

\section{Non-Maxwellian velocity distributions imply correlated constrained dynamics}
\label{app:correlated}
The broadening that we discuss in the current work falls under the broad umbrella of ``superstatistics''~\cite{beck2003superstatistics,cohen2004superstatistics}. As emphasized in the current work, the single particle velocity statistics is non Gaussian when superposing Maxwellians associated with a distribution of intensive parameters,
\begin{equation}
  P_{\beta}({\vec{v}}) = \int d\beta' \;P(\beta|\beta')\;
  \left(\frac{m\beta'}{2\pi}\right)^{3/2}
  \exp\!\left(-\frac{\beta' m v^2}{2}\right).
  \label{eq:Pbeta_v}
\end{equation}
The distributions $P$ are associated with a non-uniform temperature $\beta'$. In  Refs.~\cite{grant16,grant15,preprint6,preprint2,Nussinov2024},such distributions were not posited in an ad hoc manner. Ref.~\cite{preprint6} established, via rigorous inequalities and exactly solvable examples, that driving the system must broaden its effective energy density/effective temperature, and/or other intensive state variables that are changed as the system is driven (leading to a non-delta function like $P$). One such broadening mechanism is further derived in Appendix~\ref{rem:drive}. Non-Maxwellian single particle velocity statistics are also extremely well known in other arenas. Such a broadening also appears when local intensive variables fluctuate, as in the Landau Ginzburg description of strained supersolids and superconductors~\cite{tanwa}, in bubble nucleation~\cite{superheat}, and in plasmas~\cite{localeTplasma}. The local equilibrium ansatz of long-studied variables in generalized hydrodynamics~\cite{doyon} is of the same character.

Albeit ingenious, Maxwell’s derivation of the Maxwell-Boltzmann distribution for rotationally invariant systems {\it posits 
an absence of correlations amongst the individual velocity components}- i.e., a factorization of the velocity probability distribution into that of the individual Cartesian velocity components from rotational symmetry. As is well known, this assumption falters in many systems. The breaking of this assumption (and thus the absence of Maxwellian velocity  distributions) implies the existence of nontrivial correlations. A factorization of the probability distribution does not arise when the velocity components are not independent. The lack of Maxwellian statistics in rotationally symmetric systems mandates the appearance of nontrivial constraints/correlations amongst the velocity components. The velocities of free bosons/fermions conform to rotationally invariant Bose-Einstein and Fermi-Dirac distributions. However, these functions of the squared velocity ${\vec{v}}^2$ (derivable from geometric series in the rotationally invariant Maxwellian times the fugacity) {\it are not} products of the probability distributions of the individual velocity components $v_{a={x,y,z}}$. In an analogous way, in a relativistic system, the velocity component cannot be arbitrary and Lorentz invariance mandates that these are constrained so a factorization into the Cartesian components in void- this leads, once again, to a departure from a  classical (in this case, non-relativistic) Maxwellian distribution of velocities. Similarly, our proposed distribution is manifestly rotationally invariant (since it’s a weighted sum/integral of rotationally invariant Maxwellians, but this, again, is {\it not} a product of the probability distributions of the individual velocity components, 
\begin{eqnarray}
\label{ineqMax}
 P_{\beta}({\vec{v}}) \neq \prod_{a} P_{\beta}(v_a).    
\end{eqnarray}
Now here is the central point of the current Appendix:
Inequalities of the form of Eq. (\ref{ineqMax}) for Bose and Fermi as well as relativistic systems do, similar to those proposed in our current work for glassformers imply {\it correlated/constrained dynamics (inhibited correlated dynamics are not unexpected for glassformers)}.  We remark that non-factorizable (non-Gaussian) distributions are also found even with conventional Nose-Hoover and other thermostats that mandate a Maxwell-Boltzmann distribution for the velocity. Specifically, the self part of the van Hove function (probability distribution for particle displacements) has, for large times, a fat (exponential) tail for large displacements that cannot be factored into the respective probability distributions for displacements along the different Cartesian directions. This fat tail augmenting the Gaussian (diffusive) probability distribution for small displacements is a hallmark of dynamical heterogeneities (see also Appendix \ref{app:nongaussian}) which similarly superpose large and small displacements \cite{BB,Chauduri,Bhowmik}.


\section{Viscosity Measurement via the SLLOD Algorithm}
\label{app:sllod}

To compute the shear viscosity in the non-equilibrium steady state, we employed homogeneous shear MD using the SLLOD algorithm coupled with an iso-kinetic thermostat. This method generates planar Couette flow with a linear streaming profile while preserving periodic boundary conditions.

For a simple shear flow in the $x$-direction with velocity gradient along $y$, the streaming velocity field is
$\mathbf{u}(\mathbf{r}) = \dot{\gamma} y \,\mathbf{e}_x$,
where $\dot{\gamma}$ is the imposed shear rate. The microscopic shear stress was computed as
\begin{equation}
\sigma_{xy} = \frac{1}{V}
\left[
\sum_i m_i c_{ix} c_{iy}
+
\sum_{i<j} r_{ij,x} f_{ij,y}
\right],
\end{equation}
where $\mathbf{c}_i$ are peculiar velocities and $\mathbf{f}_{ij}$ are pair forces.
The shear viscosity was obtained from the steady state average:
$\eta = \langle \sigma_{xy} \rangle / \dot{\gamma}$.

The SLLOD equations were originally introduced to generate homogeneous shear flow in periodic systems \cite{EvansMorriss1984,Hoover1985,EvansMorrissBook}. For validation, the non-equilibrium viscosity was compared to equilibrium Green--Kubo calculations,
\begin{equation}
\eta = \frac{V}{T} \int_0^{\infty} \langle \sigma_{xy}(0)\sigma_{xy}(t) \rangle \, dt,
\end{equation}
and both approaches yielded consistent viscosity values within statistical uncertainty, confirming that the imposed shear rates remained within the linear response regime.

\section{Cooling Protocol and steady state Verification}
\label{app:methods}

\subsection{Cooling Rates and Temperature-Time Profiles}
We cooled the system to time dependent  target temperatures following an exponential quench protocol,
\begin{equation}
T_{\mathrm{target}}(t)
=
T_{\mathrm{final}}
+
\left(T_0 - T_{\mathrm{final}}\right)
e^{-t/\tau_{\mathrm{Quench}}}
\end{equation}
where \(T_0\) is the initial temperature, \(T_{\mathrm{final}}\) the final target low-temperature state, and \(\tau_{\mathrm{Quench}}\) controls the cooling rate. Small  \(\tau_{\mathrm{Quench}}\lesssim 10^2\) MD step values  correspond to rapid quenches that strongly drive the system out of equilibrium and favor glassy arrest. By contrast, larger  \(10^3 \lesssim \tau_{\mathrm{Quench}} \lesssim 10^5\) step values allow for structural relaxation and partial or full crystallization depending on thermostat noise and polydispersity. In the limit \(\tau_{\mathrm{Quench}}\to\infty\), the protocol approaches quasi-static cooling.

\subsection{steady state Verification: $F(q,t)$ as a Probe of Steady State}

The static structure factor $S(\vec{q})$ quantifies density correlations in reciprocal space and is defined as:
\begin{equation}
\begin{split}
S(\vec{q}) &= \frac{1}{N} \left\langle \sum_{j=1}^{N} \sum_{k=1}^{N}
e^{-i \vec{q} \cdot (\vec{r}_j - \vec{r}_k)} \right\rangle \\
&= \frac{1}{N} \left\langle \left| \sum_{j=1}^{N}
e^{-i \vec{q} \cdot \vec{r}_j} \right|^2 \right\rangle
\end{split}
\end{equation}
Peaks in $S(\vec{q})$ indicate preferred interparticle spacings and provide information about the system's structural ordering. The value of $q=|\vec{q}|$ at which $S(q)$ reaches its peak is the pertinent one at which the self-part of the intermediate scattering function $F(q,t)$ was computed.

\begin{figure}[tbp]
    \centering
    \includegraphics[width=1\linewidth]{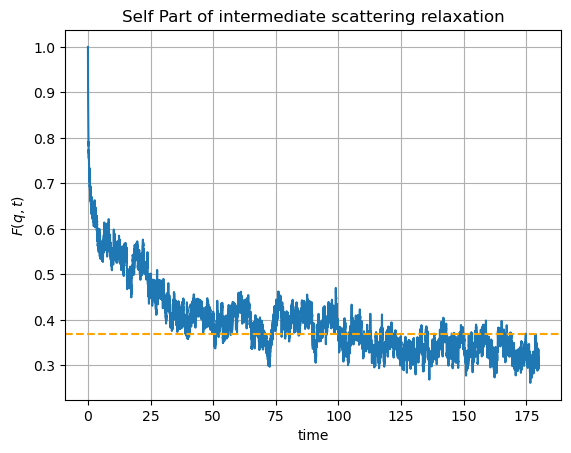}
    \caption[$F(q,t)$]{
    Representative relaxation curve of the self--intermediate scattering 
    function $F(q,t)$ at fixed wavevector $q$ for a single sample trajectory. 
    The horizontal dashed line indicates the threshold $e^{-1}$ used to define 
    the structural $\alpha$-relaxation time. 
    The vertical dashed line marks $\tau_\alpha$, determined from the condition 
    $F(q,\tau_\alpha)=e^{-1}$.}
    \label{fig:Fqt_single_sample}
\end{figure}

The corresponding time-domain relaxation of $F(\vec{q},t)$ is shown in Fig.~\ref{fig:Fqt_single_sample} for the two thermostat choices.

\subsubsection*{Stationarity Test for $F(q,t)$}

Before extracting dynamical information from the intermediate scattering
function $F(q,t)$, it is essential to verify that the system has reached
a statistically stationary state within the time window used for selecting
time origins.

\paragraph{Procedure.}
We adopt the following practical steady state test:
\begin{enumerate}
    \item Select the final $30\%$--$40\%$ of the trajectory after cooling.
    \item Divide this time window into $K = 20$ equal segments.
    \item Compute $F_k(q,t)$ independently in each segment using only
    time origins contained within that segment.
    \item Compare the resulting curves $F_k(q,t)$.
\end{enumerate}

\paragraph{Stationarity Criterion.}
We require that the relative spread across these segments satisfy 
\begin{equation}
\frac{\mathrm{std}_k\!\left[F_k(q,t)\right]}
{\left\langle F_k(q,t) \right\rangle_k}
\;\lesssim\; 3\%,
\qquad
t \in \text{$\alpha$-relaxation window}.
\end{equation}
If the block-averaged curves collapse within approximately $3\%$
throughout the decay region, the system is considered
statistically stationary and suitable for reliable extraction
of dynamical quantities such as $\tau_\alpha$. The $3\%$ margin was consistently achieved when $\kappa < 0.5$ with HTLA and TwoBath thermostats.

\bibliographystyle{apsrev4-1}
\bibliography{merged_references}

\end{document}